\documentclass[aps,prl,twocolumn,amsmath,amssymb,superscriptaddress]{revtex4}
\UseRawInputEncoding
\pdfoutput=1

\usepackage{graphicx}
\usepackage{amssymb}
\usepackage{natbib}
\usepackage{calc}
\usepackage{SIunits}
\usepackage{textcomp}

\begin{document}

\title{Preferred Spin Excitations in the Bilayer Iron-Based Superconductor CaK(Fe$_{0.96}$Ni$_{0.04}$)$_4$As$_4$ with Spin-Vortex Crystal Order}

\author{Chang Liu}
\affiliation{Beijing National Laboratory for Condensed Matter
Physics, Institute of Physics, Chinese Academy of Sciences, Beijing 100190, China}
\affiliation{School of Physical Sciences, University of Chinese Academy of Sciences, Beijing 100190, China}
\author{Philippe Bourges}
\email{philippe.bourges@cea.fr}
\affiliation{Laboratoire L\'{e}on Brillouin, CEA-CNRS, Universit\'{e} Paris-Saclay, CEA Saclay,
91191 Gif-sur-Yvette, France}
\author{Yvan Sidis}
\affiliation{Laboratoire L\'{e}on Brillouin, CEA-CNRS, Universit\'{e} Paris-Saclay, CEA Saclay,
91191 Gif-sur-Yvette, France}
\author{Tao Xie}
\affiliation{Neutron Scattering Division, Oak Ridge National Laboratory, Oak Ridge, Tennessee 37831, USA}
\author{Guanghong He}
\affiliation{International Center for Quantum Materials, School of Physics, Peking University, Beijing 100871, China}
\author{Fr\'{e}d\'{e}ric Bourdarot}
\affiliation{Universit\'{e} Grenoble Alpes, CEA, INAC, MEM MDN, F-38000 Grenoble, France}
\author{Sergey Danilkin}
\affiliation{Australian Centre for Neutron Scattering, Australian Nuclear Science and
Technology Organization, Lucas Heights NSW-2234, Australia}
\author{Haranath Ghosh}
\affiliation{Human Resources Development Section, Raja Ramanna Centre for Advanced Technology, Indore 452013, India}
\affiliation{Homi Bhabha National Institute, BARC training school complex 2nd floor, Anushakti Nagar, Mumbai 400094, India}
\author{Soumyadeep Ghosh}
\affiliation{Human Resources Development Section, Raja Ramanna Centre for Advanced Technology, Indore 452013, India}
\affiliation{Homi Bhabha National Institute, BARC training school complex 2nd floor, Anushakti Nagar, Mumbai 400094, India}
\author{Xiaoyan Ma}
\affiliation{Beijing National Laboratory for Condensed Matter
Physics, Institute of Physics, Chinese Academy of Sciences, Beijing
100190, China}
\affiliation{School of Physical Sciences, University of Chinese Academy of Sciences, Beijing 100190, China}
\author{Shiliang Li}
\affiliation{Beijing National Laboratory for Condensed Matter
Physics, Institute of Physics, Chinese Academy of Sciences, Beijing
100190, China}
\affiliation{School of Physical Sciences, University of Chinese Academy of Sciences, Beijing 100190, China}
\affiliation{Songshan Lake Materials Laboratory, Dongguan, Guangdong 523808, China}
\author{Yuan Li}
\email{yuan.li@pku.edu.cn}
\affiliation{International Center for Quantum Materials, School of Physics, Peking University, Beijing 100871, China}
\affiliation{Collaborative Innovation Center of Quantum Matter, Beijing 100871, China}
\author{Huiqian Luo}
\email{hqluo@iphy.ac.cn}
\affiliation{Beijing National Laboratory for Condensed Matter
Physics, Institute of Physics, Chinese Academy of Sciences, Beijing
100190, China}
\affiliation{Songshan Lake Materials Laboratory, Dongguan, Guangdong 523808, China}

\date{\today}

\begin{abstract}
The spin-orbit coupling (SOC) is a key to understand the magnetically driven superconductivity in iron-based superconductors, where both local and itinerant electrons are present and the orbital angular momentum is not completely quenched.  Here, we report a neutron scattering study on the bilayer compound CaK(Fe$_{0.96}$Ni$_{0.04}$)$_4$As$_4$ with superconductivity coexisting with a non-collinear spin-vortex crystal magnetic order that preserves the tetragonal symmetry of Fe-Fe plane. In the superconducting state, two spin resonance modes with odd and even $L$ symmetries due to the bilayer coupling are found similar to the undoped compound CaKFe$_4$As$_4$ but at lower energies. Polarization analysis reveals that the odd mode is $c-$axis polarized, and the low-energy spin anisotropy can persist to the paramagnetic phase at high temperature, which closely resembles other systems with in-plane collinear and $c-$axis biaxial magnetic orders. These results provide the missing piece of the puzzle on the SOC effect in iron-pnictide superconductors, and also establish a common picture of $c-$axis preferred magnetic excitations below $T_c$ regardless of the details of magnetic pattern or lattice symmetry.
\end{abstract}

\maketitle
The iron-based superconductivity emerges from a magnetic Hund's metal with strong interplay between the itinerant charge carriers on the Fermi surfaces and the local spins on Fe atoms \cite{Chen2014,Dai2012,Si2016,Georges2013}. In general, the magnetic orders in iron pnictide superconductors can be described by a unified function of the spatial variation of the Fe moments $\mathbf{m}$ at position $\mathbf{R}$: $\mathbf{m(R)}=\mathbf{M_1}\cos(\mathbf{Q_1}\cdot\mathbf{R})+\mathbf{M_2}\cos(\mathbf{Q_2}\cdot\mathbf{R})$, where $\mathbf{M_{1,2}}$ are two magnetic order parameters associated with two orthogonal wavevectors $\mathbf{Q_{1,2}}$ [\textit{e.g.}, $\mathbf{Q_{1}}=(\pi, 0)$ and $\mathbf{Q_{2}}=(0, \pi)$ ]\cite{Fernandes2016}. There are three confirmed magnetic patterns listed below \cite{Gong2018}: (1) the collinear stripe-type order with in-plane moments given by $\mathbf{M_1}\neq0$ and $\mathbf{M_2}=0$, so called as stripe spin-density wave (SSDW) \cite{Lorenzana2008,Cruz2008}; (2) the collinear biaxial order with $c-$axis polarized moments given by $\mid\mathbf{M_1}\mid=\mid\mathbf{M_2}\mid\neq0$ and $\mathbf{M_1}\parallel\mathbf{M_2}$, so called as charge-spin-density wave (CSDW) \cite{Allred2016,Bohmer2015}; (3) the non-collinear, coplanar order with in-plane moments given by $\mid\mathbf{M_1}\mid=\mid\mathbf{M_2}\mid\neq0$ and $\mathbf{M_1}\perp\mathbf{M_2}$, so called as spin-vortex crystal (SVC) phase \cite{Meier2018,Kreyssig2018}. While the latter two cases preserve the tetragonal symmetry of the lattice, the presence of the first one has to break the $C_4$ rotational symmetry by a tetragonal-to-orthorhombic structural transition, leading to an Ising-nematic electronic phase prior to the magnetic ordering \cite{Dai2015,Lu2014}.

All the three magnetic phases are proximate to each other in the mean-field phase diagram within the Ginzburg-Landau framework [Fig. 1(a)], where the magnetic ground state is determined by the signs of Landau parameters ($g$, $w$ and $\eta$) and the form of spin-orbit coupling (SOC) in actual materials \cite{Fernandes2016,Bohmer2020}. The coupling between local spins and five $3d$ Fe-orbitals is essential not only to stabilize the static magnetic order but also to modify the dynamic spin fluctuations \cite{Christensen2014,Inosov2016}. Since the spin fluctuations most likely promote the sign-reversed superconducting pairing between the hole and electron Fermi pockets (so-called $s^{\pm}-$pairing) \cite{Maier2009,Wang2013a}, the SOC will certainly show influence on the formation of superconductivity. Such effect can be traced in the anisotropy of the neutron spin resonant mode around the ordered wave vector $\mathbf{Q}$ , which is usually interpreted as an excitonic bound state from the renormalized spin excitations in the superconducting state \cite{Eschrig2006,Sidis2007,Christianson2008,Xie20181,Hong2020,Xie2021}. Interestingly, it seems that the spin resonance in many iron pnictide superconductors is preferentially polarized along $c$ axis, not only for those compounds with in-plane SSDW order [\textit{e.g.}, BaFe$_{2-x}$(Ni,Co)$_x$As$_2$, BaFe$_2$(As$_{1-x}$P$_x$)$_2$ and NaFe$_{1-x}$Co$_x$As] \cite{Lipscombe2010,Liu2012,Steffens2013,Luo2013,Zhang2014,Song2017,Hu2017,Waber2017}, but also for the underdoped (Ba,Sr)$_{1-x}$Na$_x$Fe$_2$As$_2$ with $c-$axis oriented CSDW \cite{Guo2019,Waber2019} and the optimally doped Ba$_{1-x}$K$_x$Fe$_2$As$_2$ \cite{Qureshi2014a,Zhang2013,Song2016}. This mode seems to be fully $c-$axis polarized in paramagnetic FeSe \cite{Ma2017}. Whether the resonant mode in iron-based superconductors is universally $c-$axis polarized needs to be finally investigated in the compounds with coplanar SVC order.

\begin{figure}[t]
\includegraphics[width=0.48\textwidth]{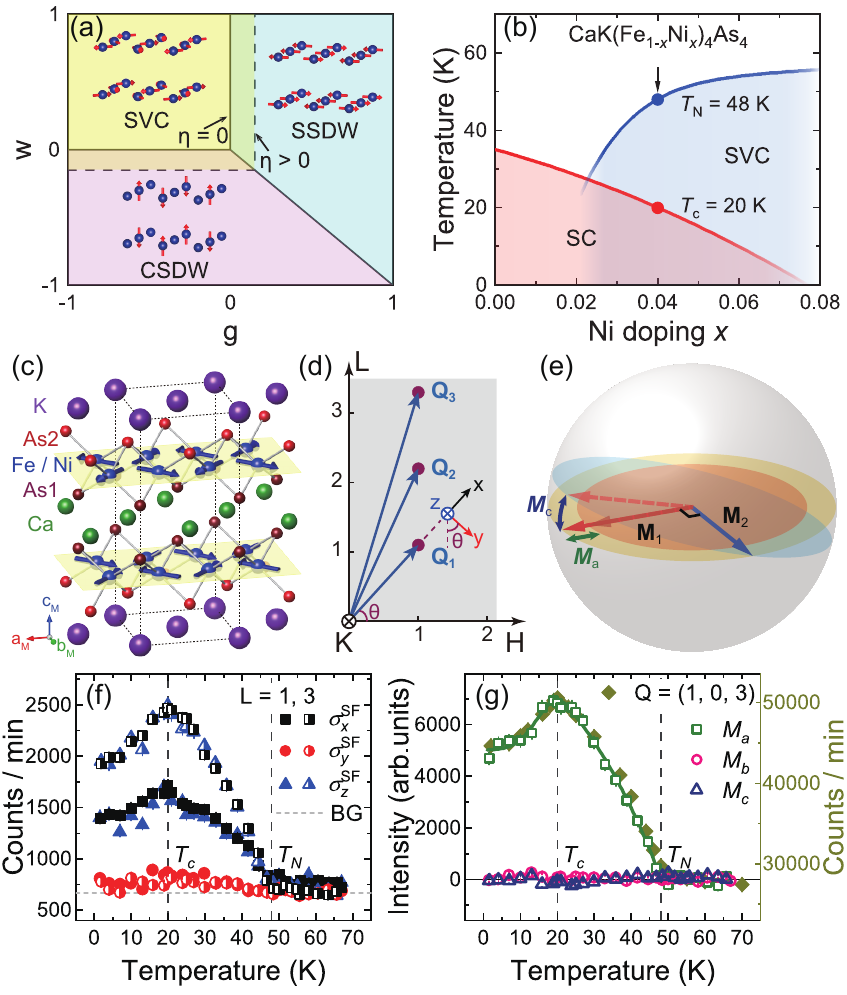}
\caption{
(a) Mean-field phase diagram of the magnetic states in iron pnictides \cite{Bohmer2020}. (b)(c) Phase diagram and SVC magnetic structure of CaK(Fe$_{1-x}$Ni$_{x}$)$_4$As$_4$ \cite{Meier2018}. The arrow marks the doping in this study.
(d) The scattering plane and the definition of spin-polarization directions in reciprocal space. (e) Schematic picture of the fluctuating moments under SVC order. Assuming $\mathbf{M_2}$ is fixed, $\mathbf{M_1}$ is allowed to fluctuate either transverse out-of-plane ($M_c$) or longitudinal in-plane ($M_a$). (f) Magnetic order parameter at $Q=(1, 0, 1)$ and (1, 0, 3) measured by polarized elastic neutron scattering. (g) Three components of static moments (open) in comparison with the unpolarized results (solid).
}
\end{figure}

Here in this Letter, we address the issue of spin anisotropies in the bilayer iron-based superconductor CaK(Fe$_{1-x}$Ni$_{x}$)$_4$As$_4$ with coexisting superconductivity and SVC magnetic order. Our unpolarized inelastic neutron scattering measurements reveal two spin resonance modes at about 7 and 15 meV with odd and even $L$ symmetries, respectively. Polarization analysis suggests that the odd mode is highly anisotropic, manifested by a strong $c-$axis component and two weakly anisotropic in-plane components ($M_c \gg M_b > M_a$). The $c-$axis preferred spin excitations already show up in the SVC phase and even continue to the paramagnetic phase until the spin anisotropy finally disappears at about $T$ = 100 K, which probably is a consequence from the SOC associated with the spontaneously broken electronic symmetries in SVC and its vestigial states. These results suggest a common feature of $c-$axis preferred spin excitations below $T_c$ regardless of their magnetic patterns or lattice symmetries in iron pnictide superconductors.

The CaK(Fe$_{1-x}$Ni$_{x}$)$_4$As$_4$ system belongs to 1144-type iron-based superconductor, where the stoichiometric compound CaKFe$_4$As$_4$ already shows a superconductivity below $T_c=35$ K without any magnetic order \cite{Iyo2016,Meier2016}. The Ni doping will suppress the superconducting transition and induce a SVC order when $x \geq 0.02$ [Fig. 1(b)]\cite{Meier2018,Kreyssig2018}. Its lattice retains the tetragonal symmetry against the chemical doping and temperature, but the translational symmetry along $c$ axis is broken due to the two inequivalent As sites below and above the Fe-Fe plane [Fig. 1(c)] \cite{Meier2017,Cui2017}. As the moments $\mathbf{M_{1,2}}$ are mutually perpendicular to preserve the tetragonal lattice symmetry, the continuous rotational symmetry is broken from $O(3)$ to $O(2)$ for the magnetism [Fig. 1(e)]. Thus a unique SOC is theoretically proposed in the form of a vector chiral order parameter $\mathbf{\varphi}=2\omega\langle\mathbf{M_1}\times\mathbf{M_2}\rangle$, giving a vestigial phase called spin-vorticity density wave (SVDW) \cite{Fernandes2016,Christensen2014,Wang2014a,Bohmer2020}. In our previous neutron scattering measurements on CaKFe$_4$As$_4$, it is found that the spin resonance splits into two nondegenerate modes with odd or even $L$ modulations induced by the broken translational symmetry along $c$ axis \cite{Xie20182}, and the low-energy odd modes are indeed $c$-axis polarized \cite{Xie2020}. Here, we focus on the Ni doped compound CaK(Fe$_{0.96}$Ni$_{0.04}$)$_4$As$_4$ with coexisting superconductivity ($T_c=20.1 \pm 0.9$ K) and SVC order ($T_N = 48.1 \pm 2.1$ K) \cite{Supplementary}. The crystals were grown by self-flux method with less than 5\% inhomogeneity of Ni doping, and co-aligned on the scattering plane $[H, 0, 0] \times [0, 0, L]$ by hydrogen-free glue \cite{Xie20182,Xie2020,Supplementary}. The magnetic unit cell ($a_M= b_M= 5.398$ \AA, $c=12.616$ \AA) was used to define the wave vector ${\bf Q}$ at ($q_x$, $q_y$, $q_z$) by $(H,K,L) = (q_xa_M/2\pi, q_yb_M/2\pi, q_zc/2\pi)$ in reciprocal lattice units (r.l.u.) \cite{Xie20182}. Unpolarized neutron scattering experiments were carried out using the Taipan spectrometer at ACNS, ANSTO, Australia, and polarized neutron scattering experiments were performed at the CEA-CRG IN22 spectrometer in ILL, Grenoble, France, using the CryoPAD system and Heusler monochromator/analyzer \cite{Berna2005}. The neutron-polarization directions were defined as $x$, $y$, $z$, with $x$ parallel to the momentum transfer $\textbf{Q}$, and $y$ and $z$ perpendicular to \textbf{Q} [Fig. 1(d)]. We focused on the spin-flip (SF) channels both before and after the sample given by three cross-sections: $\sigma_{x}^{\textrm{SF}}$, $\sigma_{y}^{\textrm{SF}}$, and $\sigma_{z}^{\textrm{SF}}$, which is only sensitive to the magnetic excitations/moments that are perpendicular to ${\bf Q}$ \cite{Lipscombe2010,Liu2012,Steffens2013,Luo2013,Qureshi2014a,Zhang2014,Song2017,Song2016,Hu2017,Zhang2013,Waber2017,Guo2019,Waber2019}. The measured intensity is related to the magnetic form factor $F(Q)$ (or structural factor $f_M(Q)$) and orientation of ordered moments, it can be also affected by the instrument resolution and higher-harmonic scattering from the monochromator \cite{Supplementary,Gong2018b,gshirane2004}. After considering the geometries of lattice and polarization at two unparallel but equivalent wavevectors, we can separate all three magnetic excitation components: $M_a$, $M_b$, and $M_c$ along the lattice axes $a_M$, $b_M$, and $c$, respectively \cite{Luo2013,Song2016,Hu2017,Xie2020,Supplementary,Xie20181}. The background from the leakage of the non-spin-flipping channel and the incoherent scattering can be also estimated \cite{Xie2020,Supplementary}.

We first present the elastic neutron scattering results to confirm the SVC order. The SVC phase is antiferromagnetically ordered along $c-$axis with magnetic peaks at ${Q}=(1, 0, L)$ ($L=0, \pm1, \pm2, \pm3, \pm4...$), where the peaks at odd $L$s are much stronger than those at even $L$s \cite{Kreyssig2018}. Fig.1(f) shows the temperature dependence of magnetic peak intensity (magnetic order parameter) at $Q=(1, 0, 1)$ and (1, 0, 3) for all three SF channels. The magnetic peak intensities concentrate in $\sigma_{x}^{\textrm{SF}}$ and $\sigma_{z}^{\textrm{SF}}$ channels, leaving a nearly flat background in $\sigma_{y}^{\textrm{SF}}$, where the competition between the magnetism and superconductivity suppresses the intensity below $T_c=20.1$ K. We have estimated all the three components of the static moments $M_a$, $M_b$, and $M_c$ as shown in Fig. 1(g). Only the in-plane component $M_a$ has non-zero value and shows same temperature dependence as unpolarized results. Therefore, the ordered moments indeed lay within the $ab$ plane where the moments $\mathbf{M_1}$ are antiferromagnetic along $\mathbf{Q_1}=(\pi, 0)$, consistent to the geometry of coplanar SVC order. The other ordered wavevector $\mathbf{Q_2}=(0, \pi)$ of $\mathbf{M_2}$ cannot be reached in this scattering plane.

\begin{figure}[t]
\includegraphics[width=0.48\textwidth]{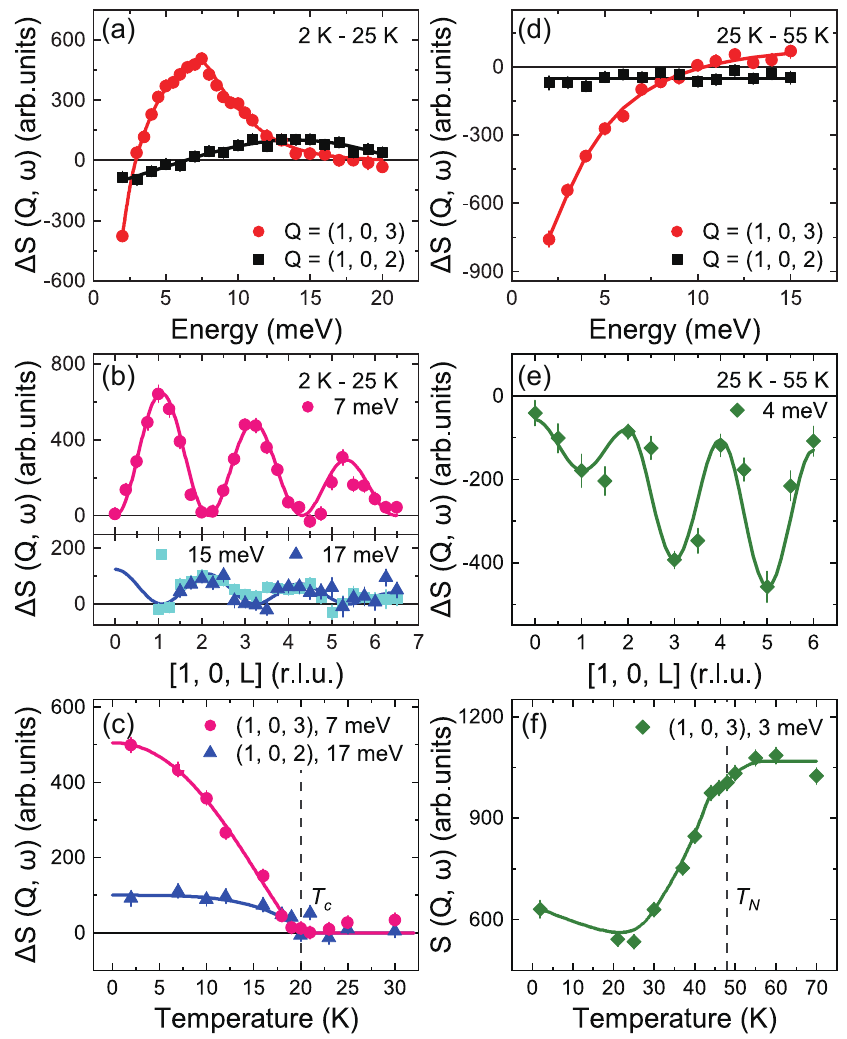}
\caption{Unpolarized neutron scattering results on the low-energy spin excitations in CaK(Fe$_{0.96}$Ni$_{0.04}$)$_4$As$_4$. (a) The odd and even spin resonant modes at $Q=(1, 0, 3)$ and (1, 0, 2), respectively. (b) $L-$modulations of two spin resonant modes at $E=7$, 15 and 17 meV. (c) Temperature dependence of the spin resonant intensity at $E=7$ and 17 meV. (d) Partially gapped spin excitations at $Q=(1, 0, 3)$ in comparison with those at $Q=$(1, 0, 2). (e) $L-$modulations of the spin gap at $E=4$ meV. (f) Temperature dependence of the intensity at $E=3$ meV and $Q=(1, 0, 3)$. All solid lines are guides to the eyes.
 }
 \end{figure}

\begin{figure}[t]
\includegraphics[width=0.48\textwidth]{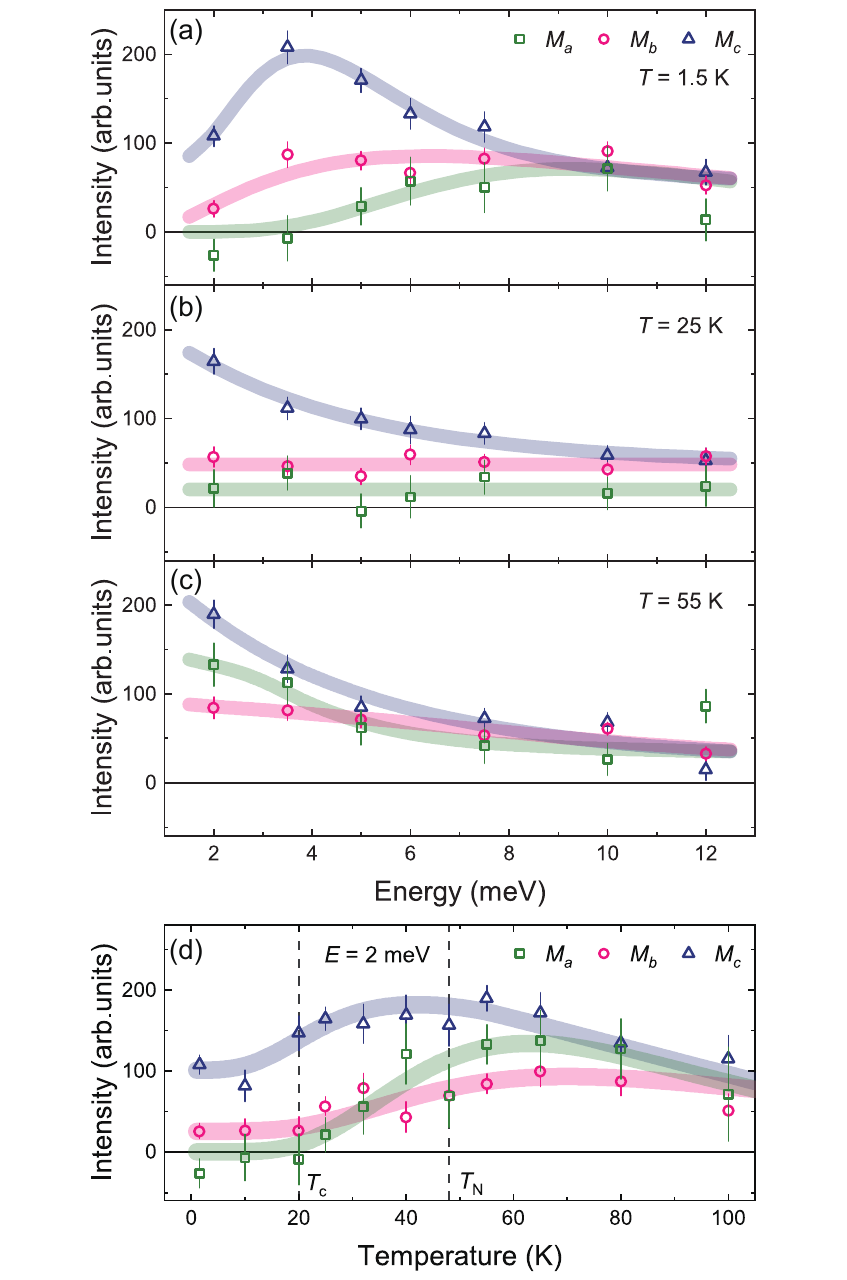}
\caption{Polarization analysis on the low-energy spin excitations in CaK(Fe$_{0.96}$Ni$_{0.04}$)$_4$As$_4$. (a) - (c) Energy dependence of $M_{a}$, $M_{b}$, and $M_{c}$ at $T=$ 1.5, 25 and 55 K deduced from three SF cross-sections: $\sigma_{x}^{\textrm{SF}}$, $\sigma_{y}^{\textrm{SF}}$, and $\sigma_{z}^{\textrm{SF}}$. (d) Temperature dependence of $M_{a}$, $M_{b}$, and $M_{c}$ at 2 meV. All bold lines are guides to the eyes.
 }
\end{figure}

The inelastic neutron scattering results with unpolarized beam are presented in Fig. 2. After subtracting the intensities in normal state (25 K) from the low temperature data in superconducting state (2 K), two spin resonant modes at $Q=(1, 0, 3)$ and (1, 0, 2) are immediately observable. While the mode at $L=3$ is centering around 7 meV, the mode at $L=2$ is around 15 meV, and both of them are broad in energy distribution [Fig. 2(a)]. The $L-$modulation of each mode follows odd symmetry $\sim\mid F(Q)\mid^2\sin^2(z\pi L)$ or even symmetry $\sim \mid F(Q)\mid^2\cos^2(z\pi L)$ respecting to the reduced distance $zc=5.849$ \AA\ ($z\approx 0.464$) within the Fe-As bilayer unit [Fig. 2(b)]. Such results are attribute to the nondegenerate spin excitations induced by the bilayer magnetic coupling, which are previously observed in the undoped compound CaKFe$_4$As$_4$ \cite{Xie20182}. By tracking the temperature dependence of intensities at $E=7$ meV, $Q=(1, 0, 3)$ and $E=17$ meV, $Q=(1, 0, 2)$, the order-parameter-like intensity gains are clearly observed below $T_c$ for both modes [Fig. 2(c)]. We also compare the intensity under SVC ordered state at $T=25$ K and paramagnetic state at $T=55$ K. It seems that the spin excitations are partially suppressed below 10 meV, giving negative values for 25 K - 55 K data with strong odd $L$ modulations [Fig. 2(d) and (e)]. After considering the effect from Bose factor, the spin gap energy is determined to be around 6 meV \cite{Supplementary}. It exhibits a transition-like behavior across $T_N$ [Fig. 2(f)]. Such spin-gapped behavior may be related to some soft excitations in the paramagnetic state, which is cleared up below $T_N$ with spectral weight condensed into the SVC Bragg peaks, and thus more obvious at $L=$odd than $L=$even \cite{Kreyssig2018}. In the SSDW systems, the spin gap is also three dimensional due to the interlayer exchange coupling $J_c$ \cite{Dai2015,Harriger2009}.

Having established the $L-$dependence of spin excitations at different temperatures, we then perform polarization analysis on the anisotropy of low-energy excitations in spin space. The raw data of the energy and temperature dependence of $\sigma_{x}^{\textrm{SF}}$, $\sigma_{y}^{\textrm{SF}}$, and $\sigma_{z}^{\textrm{SF}}$ are presented in the Supplementary Materials \cite{Supplementary}, the deduced dynamic spin components $M_a$, $M_b$, and $M_c$, are shown in Fig. 3. For the odd mode in superconducting state, there is a $c-$axis polarized component peaked around $E=4$ meV, and the in-plane excitations are also weakly anisotropic below $E=10$ meV.  At the intermediate temperature $T=25$ K, the spin resonance disappears, but the spin anisotropy still persists. When further warming up into the paramagnetic state, where the spin gap closes thus the low-energy spin excitations increase, the spin anisotropy only exists below 5 meV. Figure 3(d) shows $M_a$, $M_b$, and $M_c$ as a function of temperature at the lowest energy we can approach ($E=2$ meV). It is evident that the $c-$axis component $M_c$ is always stronger than the in-plane components $M_a$ and $M_b$, and the spin anisotropy can persist up to a very high temperature about 100 K, about twice of $T_N$. Such behavior is reminiscent of the results in BaFe$_{2-x}$Ni$_x$As$_2$, Ba$_{1-x}$K$_x$Fe$_2$As$_2$ and NaFe$_{1-x}$Co$_x$As, where the low-energy spin anisotropy can survive well above $T_N$ in the paramagnetic state \cite{Luo2013,Song2016,Song2017}.

The confirmation of $c-$axis dominated spin excitations in this work establishes a universal feature in iron pnictides that the pairing of itinerant electrons prefers to be assisted by $c-$axis spin fluctuations, no matter the magnetic pattern is in-plane collinear SSDW, $c-$axis biaxial CSDW or coplanar SVC. This could naturally explain the positive correlation between the maximum energies of SOC induced spin anisotropy and $T_c$ \cite{Song2016,Xie2020}, if the low-energy spin fluctuations indeed act as the bosonic "pairing glue" in the formation of unconventional superconductivity. The spin resonance, as a bound state of collective excitations from gapped charge carriers near the Fermi level, is formed by the spectral weight transfer from low energy to resonant energy of the spin excitations which are already present in the normal state above $T_c$. In the SSDW phase, the ordered moments are along $a$ axis and much easier to rotate perpendicularly to FeAs plane than within the plane, resulting in prominent spin fluctuations along $c$ axis ($M_c$) in the antiferromagnetic state \cite{Qureshi2012,CWang2013}. Thus $M_c$ also takes the leading contribution in the superconducting state under the SOC between itinerant electrons and local moments \cite{Luo2013,Song2017}. The in-plane spin anisotropy depends on the nematic susceptibility and the gapped intensities due to magnetic ordering \cite{Luo2013,Waber2019,Hu2017}. In the CSDW phase with $c-$axis spin reorientation, although the magnetic order already suppresses most of the longitudinal excitations ($M_c$) above $T_c$, it seems that the itinerant electrons still keep the $c-$axis "memory" in the superconducting state \cite{Guo2019,Waber2019}. Here the case of SVC phase is similar to that in SSDW phase, the in-plane longitudinal mode $M_a$ is suppressed below $T_N$ due to the magnetic ordering [Fig. 3 (b) and (c)]. On the other hand, the $c-$axis component ($M_c$) is well preserved, leaving enough spectral weight to form the spin resonance below $T_c$ [Fig. 3(a)]. The universality of $c-$axis preferred spin resonance supports the orbital-selective pairing in iron-pnictide superconductors \cite{aghosh2021,RYu2020,Chen2021}. The Cooper pairs might be favoured by particular orbital orders, which are robustly coupled to the local spins of Fe in most systems even at high temperatures far above $T_c$.

Finally, we notice that the in-plane spin excitations seem to be weakly anisotropic, such effect probably relates to the electronic nematicity involved with in-plane orbital ordering. Although the nematic order is absent in the tetragonal 1144 system, the nematic fluctuations are probed in transport measurements \cite{Bohmer2020,Zhang2018}, and possible orbital orders are proposed in the band calculations \cite{Supplementary,aghosh2021}. Moreover, the anisotropy between $M_a$ and $M_b$ can be effectively considered as the total moment on Fe sites rotates off the diagonal direction, leading to either clockwise or counterclockwise movement of the SVC-type plaquette described by the chiral order parameter of SVDW ($\mathbf{\varphi}$)\cite{Fernandes2016}. This SVDW phase is critical to affect the nematic fluctuations, as it bridges two nearly-degenerate phases (SVC and SSDW) in the mean-field phase diagram, where the parameter $\eta$ in Fig.1(a) can be viewed as a conjugate field to $\mathbf{\varphi}$ \cite{Bohmer2020,Fernandes2012}. By tuning $\eta$ from zero to finite, the degenerated region between SVC and SSDW (or between SVC and CSDW) increases, then the magnetic phase may easily transform to another under the assistant of SOC \cite{Bohmer2020}. This explains the competition between SSDW and CSDW in the hole-underdoped compounds \cite{Allred2016,Bohmer2015} as well as the additional $c-$axis moments induced by uniaxial pressure in the parent compound BaFe$_2$As$_2$ \cite{Liu2020}.

In summary, we have revealed the $L-$dependence and anisotropy of the low-energy spin excitations in the 1144-type CaK(Fe$_{0.96}$Ni$_{0.04}$)$_4$As$_4$ with superconductivity below $T_c$ = 20.1 K and a SVC order below $T_N$ = 48.1 K under the tetragonal lattice symmetry.  The bilayer coupling induced odd and even $L$ modulations of the spin resonance are observed around 7 and 15 meV, respectively. The odd mode is highly anisotropic with $c-$axis preferred magnetic excitations, and the low-energy spin anisotropy persists up to a high temperature about $T$ = 100 K, much similar to those cases with in-plane collinear or $c-$axis biaxial magnetic order. These results suggest that the $c-$axis magnetic excitations are universally preferred by the presumably orbital-selective superconducting pairing. Meanwhile, the form of magnetic order depends on material-specific symmetry characteristics in addition to SOC, leading to a rich variety of interplay between superconductivity and magnetism in the Fe-based superconductors..

The raw data are available in Ref. \cite{RawData}.

This work is supported by the National Key R\&D project of China (Grants Nos. 2020YFA0406003, 2018YFA0704200, 2018YFA0305602, and 2017YFA0302900), the National Natural Science Foundation of China (Grants Nos. 11822411, 11888101, 11874069, 11961160699, and 11674406), the Strategic Priority Research Program (B) of the CAS (Grants Nos. XDB25000000 and XDB07020300) and K.C.Wong Education Foundation (GJTD-2020-01). H. L. is grateful for the support from the Youth Innovation Promotion Association of CAS (Grant No. Y202001). Work at ORNL was supported by the U.S. Department of Energy, Office of Science, Basic Energy Sciences, Materials Science and Engineering Division. S. G. acknowledges Homi Bhaba National Institute at Raja Ramanna Centre for Advanced Technology.

\clearpage

\section{Supplementary Materials}

\begin{center}
	{\bf A. SAMPLE CHARACTERIZATION}
\end{center}

\begin{figure*}[t]
	\renewcommand\thefigure{S1}
	\includegraphics[scale = 1]{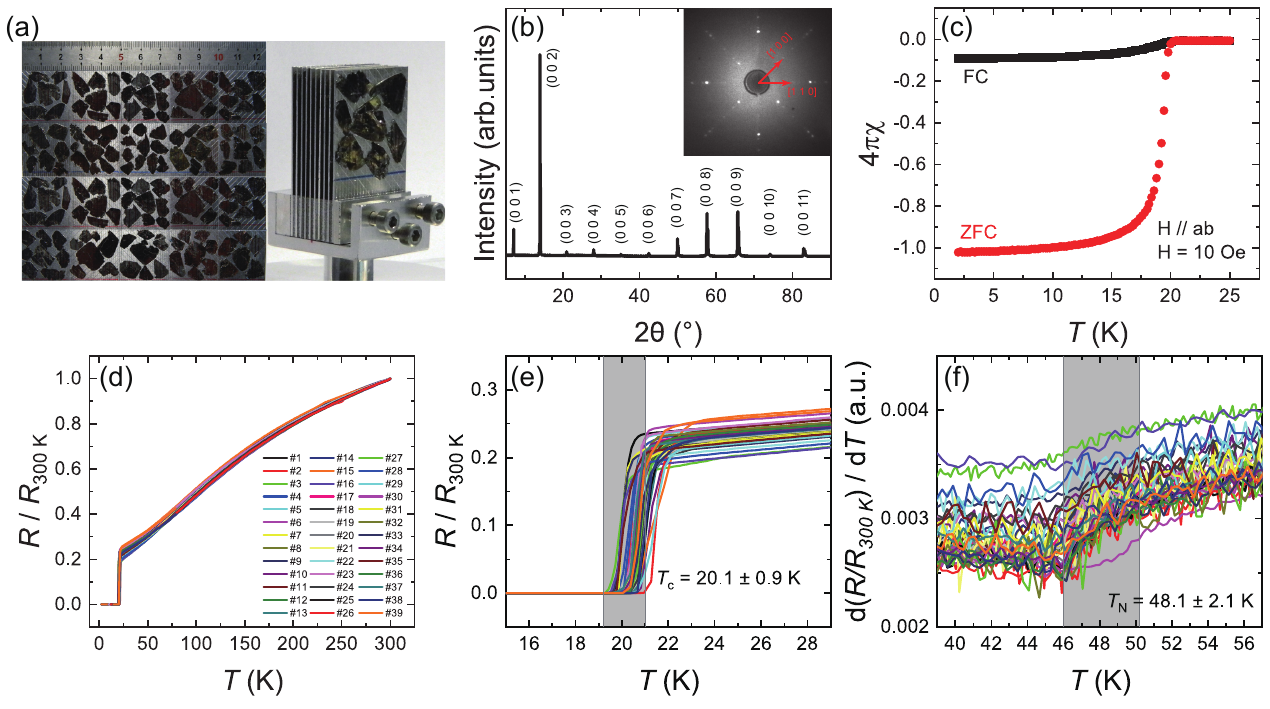}
	\caption{\label{figure_S1}(a) Photographs of the co-aligned and assembled CaK(Fe$_{0.96}$Ni$_{0.04}$)$_4$As$_4$ crystals for neutron scattering experiments. (b) X-ray diffraction pattern of CaK(Fe$_{0.96}$Ni$_{0.04}$)$_4$As$_4$ single crystal. The inset is a back-scattering Laue photograph of CaK(Fe$_{0.96}$Ni$_{0.04}$)$_4$As$_4$. The incident X-ray beam was normal to the cleaved surface. The [1, 1, 0] and [1, 0, 0] directions are marked in the tetragonal notation. (c) Temperature dependence of DC magnetic susceptibility under zero-field cooling (ZFC) and field cooling (FC). (d) Temperature dependence of the in-plane resistivity normalized by the 300 K data on 39 randomly selected samples. (e) The superconducting transitions of all measured samples, the statistics on $T_c$ defined as the starting points of zero resistance is about $20.1 \pm 0.9$ K. (f) The antiferromagnetic transitions of all measured samples. The N\'{e}el temperature $T_N$ is defined as the middle point of kink in the first-order derivative of resistivity, which is about $48.1 \pm 2.1$ K. The gray zones mark the distribution of two phase transitions.}	
\end{figure*}

We prepared high quality single crystals of CaK(Fe$_{0.96}$Ni$_{0.04}$)$_4$As$_4$ using self-flux method as our previous reports \cite{xie2018s1}. The sample photos and results of characterization are shown in Fig.~\ref{figure_S1}. About 4.426 grams of crystals were co-aligned on assembled aluminum plates by X-ray Laue camera using CYTOP hydrogen-free glue [Fig.~\ref{figure_S1}(a)]. The crystalline quality was examined by single crystal X-ray diffraction (XRD) on a SmartLab 9 kW high resolution diffraction system with Cu $K\alpha$ radiation ($\lambda$  = 1.5406 $\text{\AA}$) at room temperature ranged from $5\degree$ to $90\degree$ in reflection mode. All odd and even Bragg peaks along $c$ direction are observed in XRD measurements due to the noncentrosymmetric structure (space group: P4/mmm). The sharp $(0, 0, L)$ peaks of X-ray diffraction and fourfold symmetry of the Laue pattern in Fig.~\ref{figure_S1}(b) indicate high $c-$axis orientation of our samples. The DC magnetization measurements show sharp superconducting transitions at $T_{c}$ = 20 K and full Meissner shielding volume ($4\pi\chi\approx -1$)  [Fig.~\ref{figure_S1} (c)].  To check the chemical homogeneity of our samples, we have measured the in-plane resistivity on about 39 pieces of randomly selected crystals. All curves are nearly overlapped from 2 K to 300 K after normalizing by the room temperature resistivity [Fig.~\ref{figure_S1} (d)]. Statistically, the superconducting transition temperature ($T_c$) defined as the starting points of zero resistance, is about $20.1 \pm 0.9$ K [Fig.~\ref{figure_S1} (e)]. The magnetic transition can be also traced by the kink in the first-order derivative of resistivity, giving a N\'{e}el temperature $T_N = 48.1 \pm 2.1$ K [Fig.~\ref{figure_S1} (f)]. This means the chemical inhomogeneity of Ni doping is less than 5\%. Furthermore, the real content of Ni is about 80\% to the nominal doping ($x=0.04$) as determined by the inductively coupled plasma (ICP) analysis measurements, which is consistent with our previous experiences on the Ni doped iron-based superconductors \cite{ychen2011s,hqluo2012s,xylu2013s,rzhang2015s,txie2017s}.

\begin{center}
	{\bf B. NEUTRON SCATTERING EXPERIMENTS}
\end{center}

Unpolarized neutron scattering experiments were carried out using thermal triple-axis spectrometer Taipan at Australian Centre for Neutron Scattering, ANSTO, Australia. The final energy was fixed as $E_{f}$ = 14.87 meV, with a pyrolytic graphite filter, a double focusing monochromator and a vertical focusing analyzer. The scattering plane is $[H, 0, 0] \times [0, 0, L]$ defined using the magnetic unit cell: $a_M= b_M= 5.398$ \AA, $c=12.616$ \AA, in which the wave vector ${\bf Q}$ at ($q_x$, $q_y$, $q_z$) by $(H,K,L) = (q_xa_M/2\pi, q_yb_M/2\pi, q_zc/2\pi)$ reciprocal lattice units (r.l.u.).  The energy dependence of spin excitations were measured at $Q$ = (1, 0, 2) and (1, 0, 3) (Fig.S2). In principle, the spin resonance can be directly identified by comparing the intensities in the superconducting state ($T=2$ K) and normal state ($T=25$ K) [Fig. 2(a)], since the background from incoherent scattering, phonon excitations and multi-scattering of high order neutrons is almost the same within the probed ranges of energy and temperature. To determine the magnetic scattering at each temperature, we also measured the background at $Q=$ (0.5, 0, 2) and (0.5, 0, 3) from 2 meV to 15 meV [Fig. S2 (a) and (b)]. The neutron scattering cross section in arbitrary units [$S({\bf Q},\omega)$ ] can be obtained after subtracting the background from the raw data, assuming that most backgrounds are from $Q-$independent incoherent scattering [Fig. S2 (c) and (d)]. The local susceptibility $\chi\prime\prime({\bf Q},\omega)$ can be further obtained after correcting the Bose population factor using $\chi\prime\prime({\bf Q},\omega)=[1-\exp(-\hbar\omega/k_BT)]S({\bf Q},\omega)$, where $E=\hbar\omega$. The differences of the corresponding dynamic spin susceptibilities $\Delta\chi\prime\prime({\bf Q},\omega)$ between 2 K and 25 K are shown in Fig.S2 (e), basically similar to the results of $\Delta S({\bf Q},\omega)$ in Fig. 2(a) except for the corrections at low energies. $\Delta\chi\prime\prime({\bf Q},\omega)$ of 25 K - 55 K shows that the spin gap energy at odd $L$ is around 6 meV [Fig.S2(f)], slightly lower than that determined by $\Delta S({\bf Q},\omega)$ [Fig. 2(d)]. The Bose population factor show negligible effects on the intensities at high energy above 15 meV from 2 K to 55 K.

We carried out polarized neutron scattering experiments using the CryoPAD capability of the CEA-CRG IN22 thermal triple-axis spectrometer at the Institut Laue-Langevin, Grenoble, France. The CryoPAD system provides a strictly zero magnetic field environment for the measured sample, thus ensures the accuracy of the polarization analysis in the superconducting state \cite{LELIEVREBERNA2005s}. Polarized neutrons were produced using a focusing Heusler monochromator and analyzed using a focusing Heusler analyzer with a fixed final wave vector at $k_{f} = 2.662$ $\text{\AA}$$^{-1}$ ($E_{f} = 14.7$ meV). The scattering plane was the same as Taipan experiment, and the neutron-polarization directions were defined as $x$, $y$, $z$, with $x$ parallel to the momentum transfer $\textbf{Q}$, $y$ (in scattering plane) and $z$ (out of scattering plane) perpendicular to \textbf{Q} [Fig. 1(d)].

The effect of harmonic neutrons in the incident beam should be considered for the analysis of intensities measured in constant-$\textbf{Q}$, $E_{f}$-fixed mode with low energy\cite{neutron_book_tripleaxiss}. The flux of harmonic neutrons with wavelength $\lambda_{n} = \lambda/n$ can be fitted by the formula: $\mathrm{Flux}_{\lambda/n} (k_{i}) = (2/\pi)^{1/2}H_{n}(nk_{i})^{2+\alpha}/a^{3}\,  e^{-(nk_{i})^{2}/2a^{2}}$, in which $k_{i}$ is the incident energy of the first-order neutron. The coefficients for unpolarized neutron at IN22 spectrometer are listed below: $H_{1} = 853(84)$, $H_{2} = 255(22)$, $H_{3} = 53.9(5.7)$, $\alpha = 1.560(71)$ and $a = 2.502(30)$. Between polarized and unpolarized neutron $H_{1}$ is divided by 2 not the other $H_{n}$[Fig.S3 (a)]. It is also necessary to consider that the effectiveness of the monitor is proportional to the time spent in the monitor ($\approx 1/k$). Thus the measured intensities for the scattered beam can be corrected by multiplying by the correction factor $C= \frac{\sum_{n = 1}^{3} Flux_{\lambda/n}/n}{Flux_{\lambda}} $ [Fig.S3 (b)]. The coefficient $C$ ranges from 1.5 to 1 corresponding to the measured energy from 2 meV to 12 meV. All analyses of the intensities of polarized neutron scattering measurements shown in this paper are the results after such monitor corrections.

In the polarized neutron scattering measurements we only focused on the spin flip (SF) channels given by three cross-sections: $\sigma_{x}^{\textrm{SF}}$, $\sigma_{y}^{\textrm{SF}}$, and $\sigma_{z}^{\textrm{SF}}$ (so
called $xx$, $yy$, $zz$ channels). Because neutron SF scattering is only sensitive to the magnetic excitations/moments that are perpendicular to ${\bf Q}$ and the neutron spin-polarization directions  \cite{lipscombe2010s,liu2012s,steffens2013s,luo2013s,qureshi2014s,qureshi2012s,zhang2014s,song2016s,song2017s,hu2017s,zhang2013s}, after considering the geometries of lattice and polarization, we have  \cite{song2016s,song2017s,hu2017s}:
\begin{equation}
	\label{eq1}
	\begin{array}{l}
		\sigma_x^{\textrm{SF}}=\dfrac{R}{R+1}M_{y}+\dfrac{R}{R+1}M_{z}+BG, \\
		\\[1pt]
		\sigma_y^{\textrm{SF}}=\dfrac{1}{R+1}M_{y}+\dfrac{R}{R+1}M_{z}+BG,\\
		\\[1pt]
		\sigma_z^{\textrm{SF}}=\dfrac{R}{R+1}M_{y}+\dfrac{1}{R+1}M_{z}+BG. \\
	\end{array}
\end{equation}
Here, the spin flipping ratio $R$ represents the quality of the neutron spin polarization, defined by the leakage of nuclear Bragg peaks(should be non-spin-flip: NSF) into the SF channel: $R$ = $\sigma_{nuclear}^{\textrm{NSF}}$/$\sigma_{nuclear}^{\textrm{SF}}$, which is about 13.84 in our experiments. Then the magnitudes of magnetic excitation along the $y$ and $z$ directions $M_y$, $M_z$ and the background $BG$ can be written as \cite{qureshi2012s}:
\begin{equation}
	\label{eq2}
	\begin{array}{l}
		M_{y}=\dfrac{R+1}{R-1}\left( \sigma_x^{\textrm{SF}}-\sigma_y^{\textrm{SF}}\right), \\
		\\[1pt]
		M_{z}=\dfrac{R+1}{R-1}\left( \sigma_x^{\textrm{SF}}-\sigma_z^{\textrm{SF}}\right),  \\
		\\[1pt]
		BG = \dfrac{R}{R-1}\left( \sigma_y^{\textrm{SF}}+\sigma_z^{\textrm{SF}}\right)-\dfrac{R+1}{R-1}\sigma_x^{\textrm{SF}}. \\
	\end{array}
\end{equation}
We can determine the spin-fluctuation components $M_{\beta}$ $(\beta = a, b, c)$ along the lattice axes $a_M, b_M$ and $c$ via comparing $M_{\gamma} (\gamma = y, z)$ at two unparallel but equivalent wave vectors  ${\bf Q}_1=(1, 0, 1.1)$ and ${\bf Q}_2=(1, 0, 3.3)$ \cite{xie2018s2,xie2020s}:

\begin{equation}
	\label{eq3}
	\begin{array}{l}
	M_y({\bf Q}_1)=M_a\sin^2\theta_1+M_c\cos^2\theta_1, \\
		\\[1pt]
	eM_y({\bf Q}_2)=M_a\sin^2\theta_2+M_c\cos^2\theta_2, \\
		\\[1pt]
	M_z({\bf Q}_1) = M_b,  \\
			\\[1pt]
	eM_z({\bf Q}_2) = M_b,  \\
	\end{array}
\end{equation}
in which $\theta$ is the angle between the wavevector $\textbf{Q}$ and the $[H, 0, 0]$ direction. We have $\theta_1 = 25.2\degree$ and $\theta_2 = 54.7\degree$, respectively. Because the out of the scatting plane component $M_z$ is perpendicular to both $\textbf{Q}$ and ordered moments, it should be the same for $\textbf{Q}_{1,2}$ after considering those effects from the magnetic form factor (or structural factor) of Fe$^{2+}$ and instrument resolutions \cite{luo2013s,qureshi2014s,zhang2014s,song2016s,song2017s,hu2017s,xie2018s2,xie2020s}.  Thus we can estimate the intensity ratio factor $e$ by comparing $M_z$ at $\textbf{Q}_1$ and $\textbf{Q}_2$, and then deduce $M_a$ and $M_c$ by solving the equations above. We have used $e = 1.5$ and $2.2$ for the inelastic and elastic neutron scattering analysis, respectively. Raw data of the polarized neutron scattering measurements are shown in Fig. S4 and Fig. S5. Since the even mode of spin resonance is too weak to analyze and very likely isotropic as the case in CaKFe$_4$As$_4$, we only focus on the odd mode with local maximum at $L=$ 1.1 and 3.3 \cite{xie2020s}. The deduced $M_{\beta}$ $(\beta = a, b, c)$ are shown in Fig. 3, and the dynamic spin susceptibilities $\chi_{\beta}\prime\prime$ $(\beta = a, b, c)$ are shown in Fig. S6. We have to admit that the different estimation of $e$ would possibly give some changes on $M_{\beta}$ $(\beta = a, b, c)$ especially for the weak differences between $M_a$ and $M_b$. However, based on our previous experience \cite{luo2013s,hu2017s,xie2018s2,xie2020s}, $e$ is mostly about 1.5 $\sim$ 1.8. Although larger $e$ gives slightly smaller $M_c$ and larger $M_a$, this won't change the conclusion about $c-$axis preferred spin excitations.

In Fig. S4, both spin resonance in the superconducting state and partially opened spin gap in the SVC state are clearly observed when comparing the intensities at different temperatures, where the peak energy of the resonance may be slightly lower than those in the unpolarized experiment due to different resolutions and monitor corrections of two spectrometers.  The deduced backgrounds within low-energy region are basically flat for all temperatures at $L=$ 1.1 and 3.3. Although at the first glance to the data of $L=3.3$ at $T=1.5$ and 25 K may suggest isotropic signals with $\sigma_{y}^{\textrm{SF}}\thickapprox\sigma_{z}^{\textrm{SF}}$ [Fig. S4(d) and (e)], the spin excitations are definitely anisotropic below 10 meV for $L=1.1$ [Fig. S4(a) and (b)]. When warming up to the paramagnetic state at $T=55$ K, the spin anisotropy only exists below 5 meV [Fig. S4(c) and (f)].  The temperature dependence of the low-energy spin anisotropy at $E=2$ meV are presented in Fig. S5 both for $L=1.1$ and 3.3, showing clear differences between $\sigma_{y}^{\textrm{SF}}$ and $\sigma_{z}^{\textrm{SF}}$ up to $T=100$ K. The results of local susceptibilities $\chi_{\beta}\prime\prime$ $(\beta = a, b, c)$ in Fig. S6 are similar to the $M_{\beta}$ $(\beta = a, b, c)$  in Fig. 3.

\begin{center}
	{\bf C. DENSITY-FUNCTIONAL-THEORY CALCULATION}
\end{center}

To understand the orbital contributions of the electronic band and Fermi surfaces, we have performed systematic electronic structure calculations using Density-Functional-Theory (DFT) based first principles method for both the undoped CaKFe$_4$As$_4$ and Ni-doped CaK(Fe$_{0.96}$Ni$_{0.04}$)$_4$As$_4$ compounds. We used plane wave pseudo potential method as implemented in the QUANTUM ESPRESSO package \cite{giannozzi2017s} and treated electronic exchange correlation within the generalized gradient approximation (GGA) using Perdew-Burke-Enzerhof (PBE) functional \cite{perdew1996s}. The Ultrasoft pseudopotentials were taken from Psilibrary database. Virtual Crystal Approximation (VCA) was employed to dope the system  \cite{bellaiche2000s}, namely, the actual atoms were replaced by fictitious virtual atoms in same atomic sites. The Kohn-Sham orbitals were expanded in plane wave basis set and kinetic energy cutoff is set to 45 Ry for undoped compound and 55 Ry for Ni-doped compound, respectively. The charge density cutoff was ranging between 360 - 440 Ry for different compounds. We used Monkhorst-Pack scheme for Brillouin zone (BZ) sampling in k-space and Broyden-Fletcher-Goldfrab-Shanno (BFGS) geometry optimization scheme to yield optimized structures \cite{monkhorst1976s,pack1977s,fletcher1970s,goldfarb1970s,shanno1970s}. Fermi surface integration was executed with the Gaussian smearing technique with smearing parameter about 0.01 Ry. Orbital character of the Fermi surfaces were determined using Fermisurfer visualization tool \cite{kawamura2019s}. The DFT calculation results are summarized in Fig. S7 and Fig. S8. Similar results on CaKFe$_4$As$_4$, CaRbFe$_4$As$_4$  and CaCsFe$_4$As$_4$ have been published elsewhere \cite{ghosh2021s}, here we only discuss for CaKFe$_4$As$_4$ and CaK(Fe$_{0.96}$Ni$_{0.04}$)$_4$As$_4$.

As shown in Fig.~ S7, the Fermi surfaces of 1144-type compounds consist of six hole-like bands around the zone centre $\Gamma$-point and four electron-like bands around the zone corner $M$-point, each band is of multi-orbital in nature, which is quite similar to other iron-based superconducting families. Besides the five orbitals of Fe-3d ($d_{xz}$, $d_{yz}$, $d_{xy}$, $d_{x^2-y^2}$, $d_{z^2}$), the As-$4p_z$ orbital also contributes to the density of states (DOS) at the Fermi level \cite{ghosh2021s,lochner2017s}. However, each orbital has distinct contribution to each Fermi sheet, as their partial DOS at Fermi level changes from one to another. It is argued that the 1144-families are simultaneously orbital selective self doped systems. When replacing K by Rb and Cs, the different chemical potential of different orbital derived electron/hole bands causes crossing of the Fermi level for some bands. Further orbital selective Lifshitz transition of the Fermi surfaces are also predicted by DFT calculations \cite{ghosh2021s}. Here, we focus on the detailed contribution from each Fe-3d orbital on each Fermi sheet in CaK(Fe$_{1-x}$Ni$_{x}$)$_4$As$_4$, where the colour scheme from blue to red means gradually increasing contribution of respective orbitals. The outermost hole-like Fermi pocket is mostly contributed by $d_{yz}$ and $d_{z^2}$ orbitals, and partially by $d_{xz}$ orbital, the second outermost hole-like Fermi pocket is mainly derived from $d_{xz}$, $d_{yz}$ and $d_{z^2}$ orbitals, and the two innermost hole-like Fermi pockets are mostly contributed by $d_{x^2-y^2}$ and $d_{z^2}$ orbitals. The electron-like Fermi pockets are also orbital dependent with significant amount of $d_{xz}$ and $d_{yz}$ as well as partial $d_{x^2-y^2}$ characters. Basically, the orbital dependence of Fermi surfaces are quite similar between two compounds, where the difference is the proportion between $d_{xz}$ and $d_{yz}$ orbitals in the hole-like Fermi pockets. In CaK(Fe$_{0.96}$Ni$_{0.04}$)$_4$As$_4$, it seems more contributions are from $d_{xz}$ and less from $d_{yz}$ than those in CaKFe$_4$As$_4$ (Fig. S7 (a), (b), (f), (g)). Similar case may happen on the electron-like Fermi pockets, too. In both compounds, the occupation of $d_{xz}$ and $d_{yz}$ orbitals is different on the same Fermi pocket. This means the in-plane orbital ordering may exist in this system. In Fig. S8 we present detailed band calculation results for $d_{xz}$ and $d_{yz}$ orbitals especially for $M_x$ and $M_y$ points. Clearly, there is a splitting between $d_{xz}$ and $d_{yz}$ orbitals, where $\Delta E=E(d_{xz})-E(d_{yz})=14$ meV for undoped compound and $\Delta E=60$ meV for 4\% Ni doped compound, respectively. More interestingly, such orbital ordering changes sign between $M_x$ and $M_y$ points in both compounds, namely, $\Delta E$ is positive at $M_y$ but negative at $M_x$. Therefore, the in-plane orbital ordering could be enhanced by electron doping from Ni and likely to have Fermi surface dependence, accompanying by the emergence of spin-vortex order.

\newpage

\begin{figure*}[t]
	\renewcommand\thefigure{S2}
	\includegraphics[scale = 1]{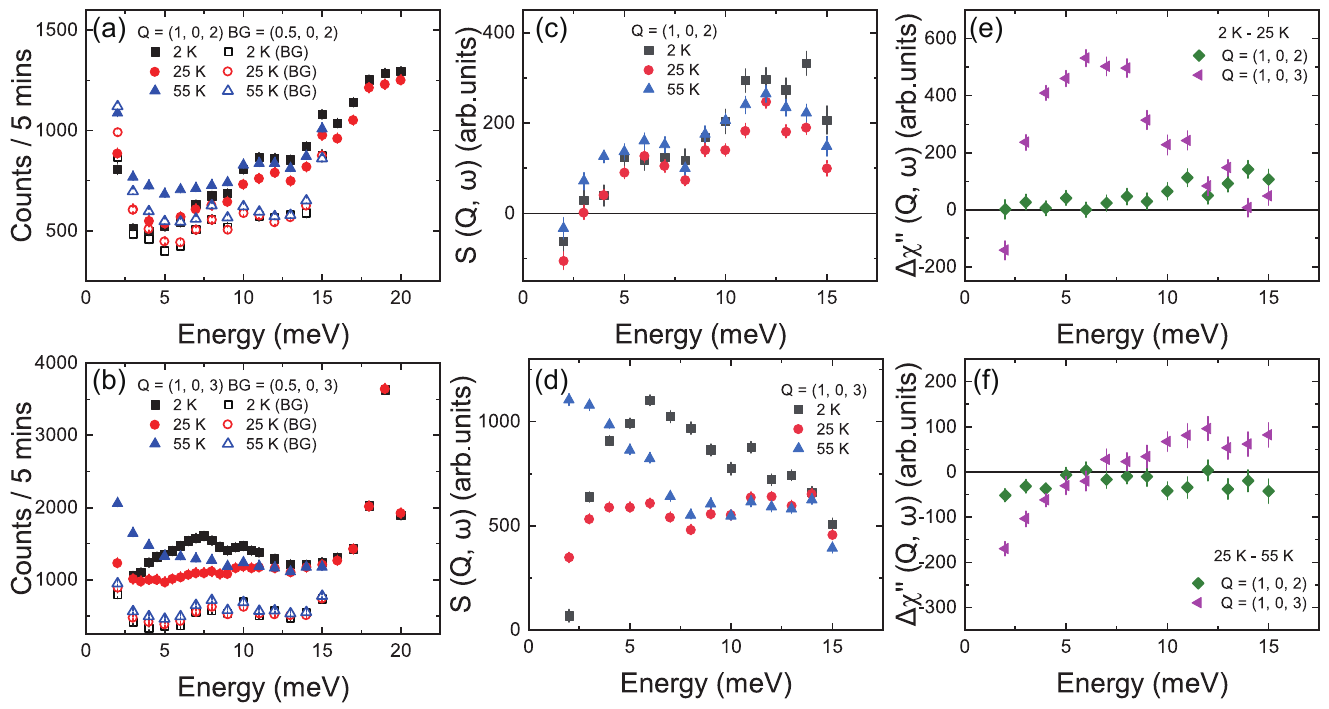}
	\caption{\label{figure_S2} Unpolarized neutron scattering data on the low-energy spin excitations in CaK(Fe$_{0.96}$Ni$_{0.04}$)$_4$As$_4$. (a) and (b) Raw data of energy scans at ${\bf Q}$ = (1, 0, 2) with background [${\bf Q}$ = (0.5, 0, 2)] and ${\bf Q}$ = (1, 0, 3) with background [${\bf Q}$ = (0.5, 0, 3)] at 1.5 K, 25 K and 55 K. (c) and (d) $S({\bf Q},\omega)$ at ${\bf Q}$ = (1, 0, 2) and ${\bf Q}$ = (1, 0, 3) obtained by subtracting the background. (e) and (f) $\Delta\chi^{''}({\bf Q},\omega)$ at ${\bf Q}$ = (1, 0, 2) and ${\bf Q}$ = (1, 0, 3) for 2 K - 25 K and 25 K - 55 K.}	
\end{figure*}

\begin{figure*}[t]
	\renewcommand\thefigure{S3}
	\includegraphics[scale = 1]{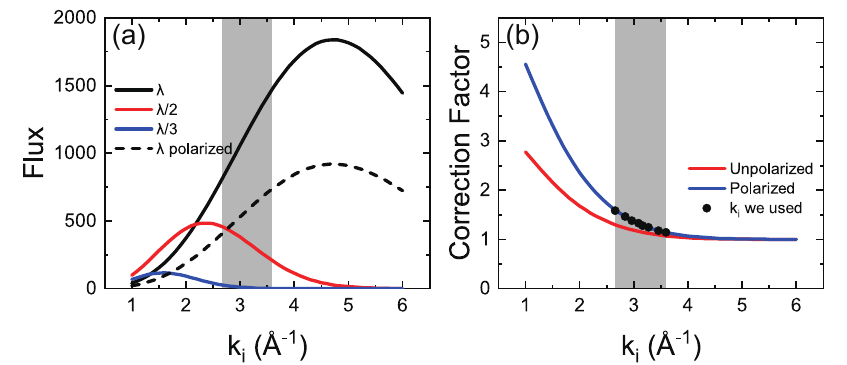}
	\caption{\label{figure_S3}(a) Flux for harmonic contents of incident neutron beam at IN22 spectrometer. The black, red and blue solid lines represent flux of the first-, second- and third-order unpolarized neutron, respectively, and the black dashed line represents the first-order polarized neutron. (b) Correction factor for fixed-$E_{f}$ scans. The black dots and shadow zone mark the $k_{i}$ we used in the polarized neutron scattering measurements. }	
\end{figure*}

\begin{figure*}[t]
	\renewcommand\thefigure{S4}
	\includegraphics[scale = 1]{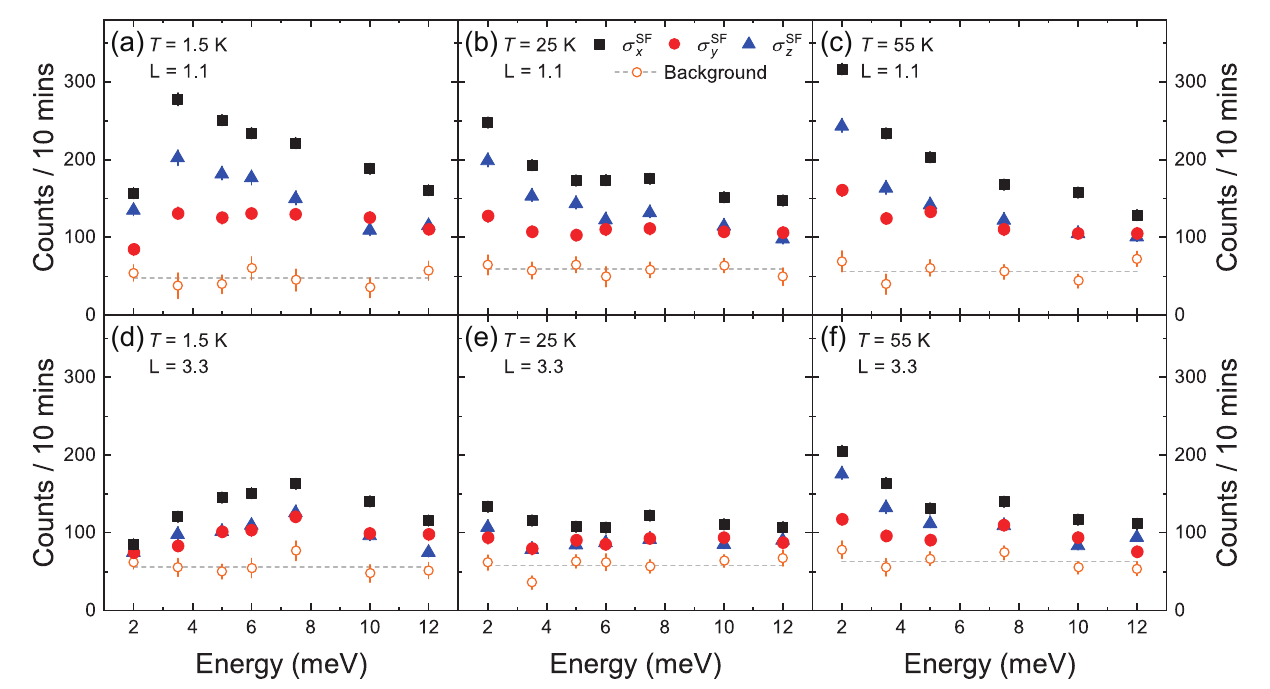}
	\caption{\label{figure_S4}Polarized neutron scattering results on the low-energy spin excitations in CaK(Fe$_{0.96}$Ni$_{0.04}$)$_4$As$_4$ at $T=$ 1.5, 25 and 55 K. The solid symbols are raw data of $\sigma_{x}^{\textrm{SF}}$, $\sigma_{y}^{\textrm{SF}}$ and $\sigma_{z}^{\textrm{SF}}$ at ${\bf Q}=(1, 0, L)$ with $L = 1.1$ and 3.3, and the open circles are the estimated backgrounds.}	
\end{figure*}

\begin{figure*}[t]
	\renewcommand\thefigure{S5}
	\includegraphics[scale = 1]{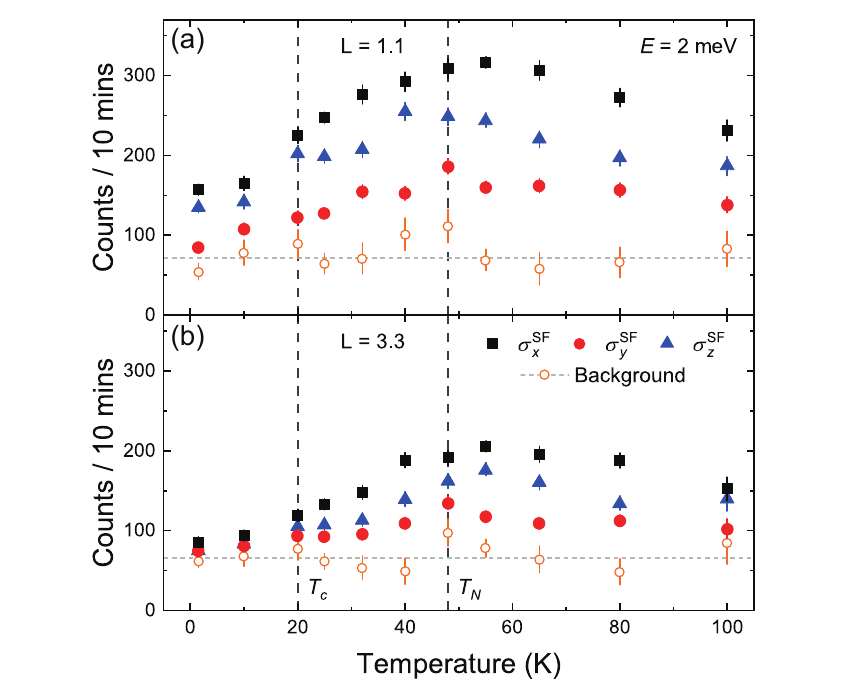}
	\caption{\label{figure_S5}Temperature dependence of the $\sigma_{x}^{\textrm{SF}}$, $\sigma_{y}^{\textrm{SF}}$, and $\sigma_{z}^{\textrm{SF}}$ at $E=2$ meV for (a) ${\bf Q}=(1, 0, 1.1)$ and (b) ${\bf Q}=(1, 0, 3.3)$, where the open circles are estimated backgrounds and the horizontal dashed lines mark their average value. }	
\end{figure*}

\begin{figure*}[t]
	\renewcommand\thefigure{S6}
	\includegraphics[scale = 1.2]{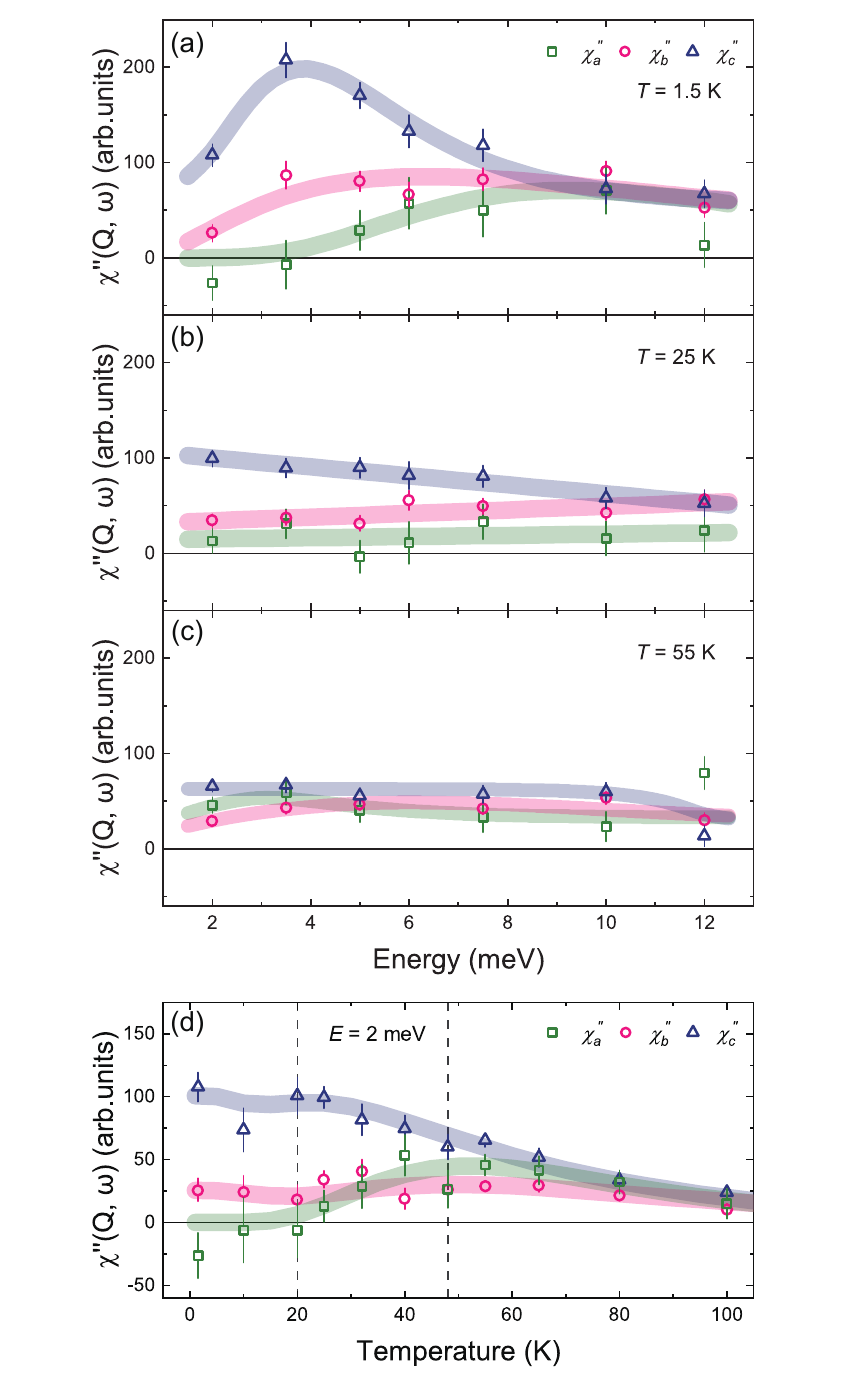}
	\caption{\label{figure_S6} Polarized neutron scattering analyses on the low-energy spin excitations in CaK(Fe$_{0.96}$Ni$_{0.04}$)$_4$As$_4$. (a) - (c) Energy dependence of $\chi_{a}^{''}$, $\chi_{b}^{''}$, and $\chi_{c}^{''}$ at $T=$ 1.5, 25 and 55 K. (d) Temperature dependence of $\chi_{a}^{''}$, $\chi_{b}^{''}$, and $\chi_{c}^{''}$ at 2 meV. All bold lines are guides to the eyes.
}
\end{figure*}

\begin{figure*}[t]
	\renewcommand\thefigure{S7}
	\includegraphics[scale = 0.75]{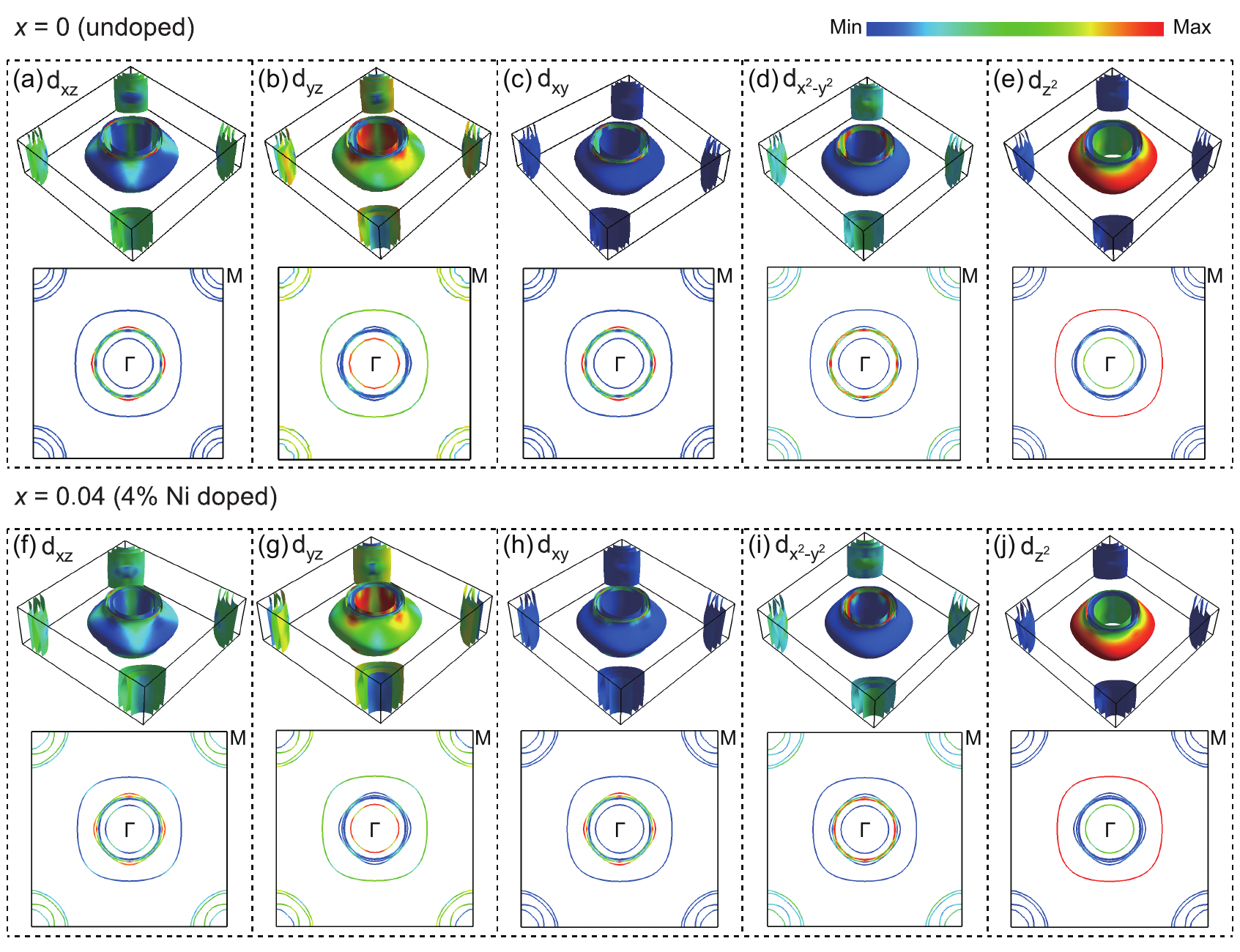}
	\caption{\label{figure_S7}DFT calculation results for the occupation of all five orbits ($d_{xz}$, $d_{yz}$, $d_{xy}$, $d_{x^2-y^2}$, $d_{z^2}$) on each Fermi sheet. (a)-(e) DFT results for undoped compound CaKFe$_4$As$_4$. (f)-(j) DFT results for Ni doped compound CaK(Fe$_{0.96}$Ni$_{0.04}$)$_4$As$_4$.}	
\end{figure*}

\begin{figure*}[t]
	\renewcommand\thefigure{S8}
	\includegraphics[scale = 0.9]{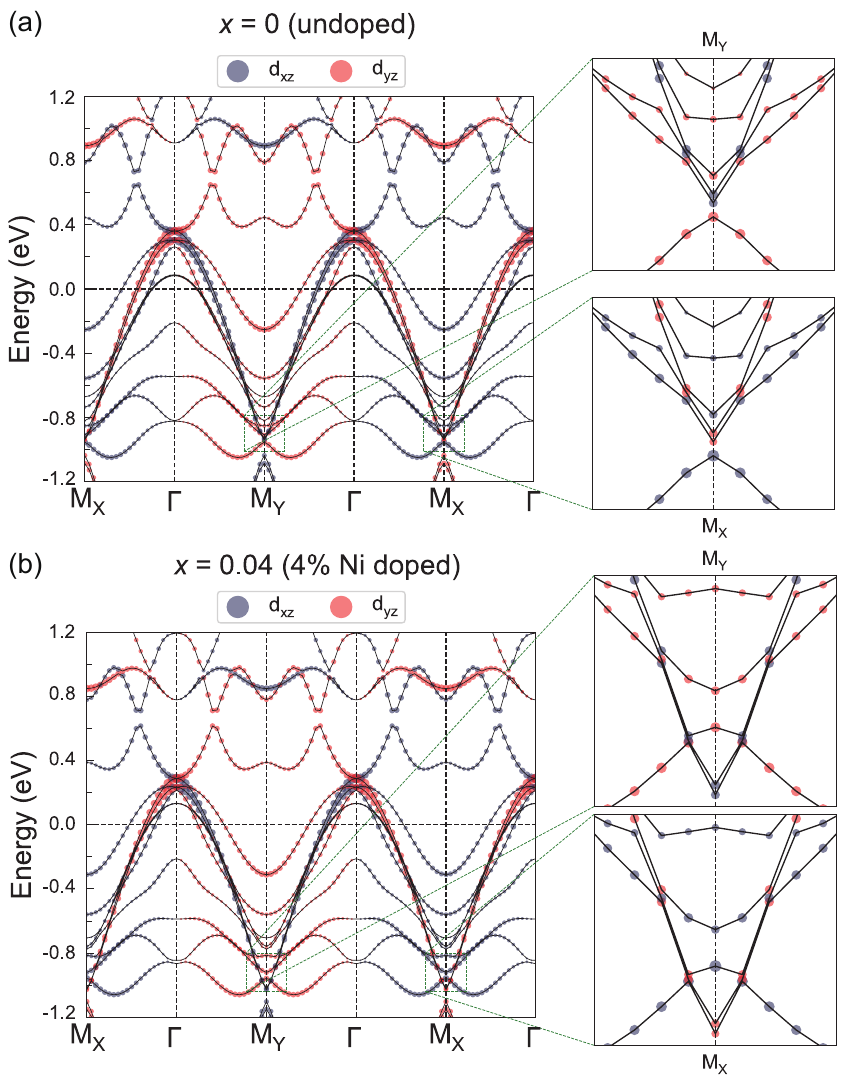}
	\caption{\label{figure_S8}DFT calculation results for (a) undoped compound CaKFe$_4$As$_4$ and (b) Ni doped compound CaK(Fe$_{0.96}$Ni$_{0.04}$)$_4$As$_4$, where the occupation of $d_{xz}$ and $d_{yz}$ orbits are marked by different color dots. The right insets are zoom in results at $M_x$ and $M_y$ points in the Brillouin zone.}	
\end{figure*}


\begin{thebibliography}{}
\bibitem{Chen2014} X. Chen, P. Dai, D. Feng, T. Xiang, and F. Zhang, Nation. Sci. Rev. {\bf 1}, 371 (2014).
\bibitem{Dai2012} P. Dai, J. Hu, and E. Dagotto, Nat. Phys. {\bf 8}, 709 (2012).
\bibitem{Si2016} Q. Si, R. Yu, and E. Abrahams, Nat. Rev. Mater. {\bf 1}, 16017(2016).
\bibitem{Georges2013} A. Georges, L. de'Medici, and J. Mravlje, Annu. Rev. Condens. Matter Phys. {\bf 4}, 137 (2013).
\bibitem{Fernandes2016} R. M. Fernandes, S. A. Kivelson, and E. Berg, Phys. Rev. B {\bf 93}, 014511 (2016).
\bibitem{Gong2018} D. Gong and H. Luo, Acta Phys. Sin. {\bf 67}(20), 207407(2018).
\bibitem{Lorenzana2008} J. Lorenzana, G. Seibold, C. Ortix, and M. Grilli, Phys. Rev. Lett. {\bf 101}, 186402 (2008).
\bibitem{Cruz2008} C. de la Cruz, Q. Huang, J. W. Lynn, J. Li, W. Ratcliff II, J. L. Zarestky, H. A. Mook, G. F. Chen, J. L. Luo, N. L. Wang and P. Dai, Nature (London) {\bf 453}, 899 (2008).
\bibitem{Allred2016} J. M. Allred, K. M. Taddei, D. E. Bugaris, M. J. Krogstad, S. H. Lapidus, D. Y. Chung, H. Claus, M.G. Kanatzidis, D. E. Brown, J. Kang, R. M. Fernandes, I. Eremin, S. Rosenkranz, O. Chmaissem, and R. Osborn, Nat. Phys. {\bf 12}, 493 (2016).
\bibitem{Bohmer2015} A. E. B\"{o}hmer, F. Hardy, L. Wang, T. Wolf, P. Schweiss, C. Meingast, Nat. Commun. {\bf 6}, 7911 (2015).
\bibitem{Meier2018} W. R. Meier, Q.-P. Ding, A. Kreyssig, S. L. Budko, A. Sapkota, K. Kothapalli, V. Borisov, R. Valent¨ª, C. D. Batista, P. P. Orth, R. M. Fernandes, A. I. Goldman, Y. Furukawa, A. E. B\"{o}hmer, and P. C. Canfield, npj Quantum Mater. {\bf 3}, 5 (2018).
\bibitem{Kreyssig2018} A. Kreyssig, J. M. Wilde, A. E. B\"{o}hmer, W. Tian, W. R. Meier, Bing Li, B. G. Ueland, Mingyu Xu, S. L. Budko, P. C. Canfield, R. J. McQueeney, and A. I. Goldman, Phys. Rev. B {\bf 97}, 224521 (2018).
\bibitem{Dai2015} P. Dai, Rev. Mod. Phys. {\bf 87}, 855 (2015).
\bibitem{Lu2014} X. Lu, J. T. Park, R. Zhang, H. Luo, Andriy. H. Nevidomskyy, Q. Si, and P. Dai, Science {\bf 345}, 657(2014).
\bibitem{Bohmer2020} A. E. B\"{o}hmer, F. Chen, W. R. Meier, M. Xu, G. Drachuck, M. Merz, P. W. Wiecki, S. L. Budko, V. Borisov, R. Valent¨ª, M. H. Christensen, R. M. Fernandes, C. Meingast, P. C. Canfield, pre-print on arXiv: 2011.13207.
\bibitem{Christensen2014} M. H. Christensen, J. Kang, R. M. Fernandes, Phys. Rev. B {\bf 100}, 014512 (2019).
\bibitem{Inosov2016} D. S. Inosov, C. R. Phys. {\bf 17}, 60 (2016).
\bibitem{Maier2009} T. A. Maier, S. Graser, D. J. Scalapino, and P. Hirschfeld, Phys. Rev. B {\bf 79}, 134520 (2009).
\bibitem{Wang2013a} Z. Wang, H. Yang, D. Fang, B. Shen, Q. -H. Wang, L. Shan, C. Zhang, P. Dai, and H. -H. Wen, Nat. Phys. {\bf 9}, 42 (2013).
\bibitem{Eschrig2006}  M. Eschrig, Adv. Phys. {\bf 55}, 47 (2006).
\bibitem{Sidis2007} Y. Sidis, S. Pailh\`{e}s, V. Hinkov, B. Fauqu\'{e}, C. Ulrich, L. Capogna, A. Ivanov, L. -P. Regnault, B. Keimer, and P. Bourges,  C. R. Phys. {\bf 8}, 745 (2007).
\bibitem{Christianson2008} A. D. Christianson, E. A. Goremychkin, R. Osborn, S. Rosenkranz, M. D. Lumsden, C. D. Malliakas, I. S. Todorov, H. Claus, D. Y. Chung, M. G. Kanatzidis, R. I. Bewley, and  T. Guidi, Nature (London) {\bf 456}, 930 (2008).
\bibitem{Xie20181} T. Xie, D. Gong, H. Ghosh, A. Ghosh, M. Soda, T. Masuda, S. Itoh, F. Bourdarot, L.-P. Regnault, S. Danilkin, S. Li, and H. Luo, Phys. Rev. Lett. {\bf 120}, 137001 (2018).
\bibitem{Hong2020} W. Hong, L. Song, B. Liu, Z. Li, Z. Zeng, Y. Li, D. Wu, Q. Sui, T. Xie, S. Danilkin, H. Ghosh, A. Ghosh, J. Hu, L. Zhao, X. Zhou, X. Qiu, S. Li, and H. Luo, Phys. Rev. Lett. {\bf 125}, 117002 (2020).
\bibitem{Xie2021} T. Xie, C. Liu, T. Fennell, U. Stuhr, S. Li, and H. Luo, Chin. Phys. B {\bf 30},127402 (2021).
\bibitem{Lipscombe2010} O. J. Lipscombe, L. W. Harriger, P. G. Freeman, M. Enderle, C. Zhang, M. Wang, T. Egami, J. Hu, T. Xiang, M. R. Norman, and P. Dai, Phys. Rev. B {\bf 82}, 064515 (2010).
\bibitem{Liu2012} M. Liu, C. Lester, J. Kulda, X. Lu, H. Luo, M. Wang, S. M. Hayden, and P. Dai, Phys. Rev. B {\bf 85}, 214516 (2012).
\bibitem{Steffens2013} P. Steffens, C. H. Lee, N. Qureshi, K. Kihou, A. Iyo, H. Eisaki, and M. Braden, Phys. Rev. Lett. {\bf 110}, 137001 (2013).
\bibitem{Luo2013} H. Luo, M. Wang, C. Zhang, X. Lu, L.-P. Regnault, R. Zhang, S. Li, J. Hu, and P. Dai, Phys. Rev. Lett. {\bf 111}, 107006 (2013).
\bibitem{Zhang2014} C. Zhang,Y. Song, L.-P. Regnault, Y. Su, M. Enderle, J. Kulda, G. Tan, Z. C. Sims, T. Egami, Q. Si, and P. Dai, Phys. Rev. B {\bf 90}, 140502(R) (2014).
\bibitem{Song2017} Y. Song, W. Wang, C. Zhang, Y. Gu, X. Lu, G. Tan, Y. Su, F. Bourdarot, A. D. Christianson, S. Li, and P. Dai, Phys. Rev. B {\bf 96}, 184512 (2017).
\bibitem{Hu2017} D. Hu, W. Zhang, Y. Wei, B. Roessli, M. Skoulatos, L.-P. Regnault, G. Chen, Y. Song, H. Luo, S. Li, and P. Dai, Phys. Rev. B {\bf 96}, 180503(R) (2017).
\bibitem{Waber2017}  F. Wa{\ss}er, C. H. Lee, K. Kihou, P. Steffens, K. Schmalzl, N. Qureshi, and M. Braden, Sci. Rep. {\bf 7}, 10307 (2017).
\bibitem{Guo2019} J. Guo, L. Yue, K. Iida, K. Kamazawa, L. Chen, T. Han, Y. Zhang, and Y. Li, Phys. Rev. Lett. {\bf 122}, 017001 (2019).
\bibitem{Waber2019}  F. Wa{\ss}er, J. T. Park, S. Aswartham, S. Wurmehl, Y. Sidis, P. Steffens, K. Schmalzl, Be. B¨¹chner, and M. Braden, npj Quantum Mater. {\bf 4}, 59 (2019).
\bibitem{Zhang2013} C. Zhang, M. Liu, Y. Su, L.-P. Regnault, M. Wang, G. Tan, Th. Br\"{u}ckel, T. Egami, and P. Dai, Phys. Rev. B {\bf 87}, 081101(R) (2013).
\bibitem{Qureshi2014a} N. Qureshi, C. H. Lee, K. Kihou, K. Schmalzl, P. Steffens, and M. Braden, Phys. Rev. B {\bf 90}, 100502(R) (2014).
\bibitem{Song2016} Y. Song, H. Man, R. Zhang, X. Lu, C. Zhang, M. Wang, G. Tan, L.-P. Regnault, Y. Su, J. Kang, R. M. Fernandes, and P. Dai, Phys. Rev. B {\bf 94}, 214516 (2016).
\bibitem{Ma2017} M. Ma, P. Bourges, Y. Sidis, Y. Xu, S. Li, B. Hu, J. Li, F. Wang, and Y. Li, Phys. Rev. X {\bf 7}, 021025 (2017).
\bibitem{Iyo2016} A. Iyo, K. Kawashima, T. Kinjo, T. Nishio, S. Ishida, H. Fujihisa, Y. Gotoh, K. Kihou, H. Eisaki, and Y. Yoshida, J. Am. Chem. Soc. {\bf 138}, 3410 (2016).
\bibitem{Meier2016} W. R. Meier, T. Kong, U. S. Kaluarachchi, V. Taufour, N. H. Jo, G. Drachuck, A. E. B\"{o}hmer, S. M. Saunders, A. Sapkota, A. Kreyssig, M. A. Tanatar, R. Prozorov, A. I. Goldman, F. F. Balakirev, A. Gurevich, S. L. Bud'ko, and P. C. Canfield, Phys. Rev. B {\bf 94}, 064501 (2016).
\bibitem{Meier2017} W. R. Meier, T. Kong, S. L. Bud'ko, and P. C. Canfield, Phys. Rev. Mater. {\bf 1}, 013401 (2017).
\bibitem{Cui2017} J. Cui, Q.-P. Ding, W. R. Meier, A. E. B\"{o}hmer, T. Kong, V. Borisov, Y. Lee, S. L. Bud'ko, R. Valent\'{\i}, P. C. Canfield, and Y. Furukawa, Phys. Rev. B {\bf 96}, 104512 (2017).
\bibitem{Wang2014a} X. Wang and R. M. Fernandes, Phys. Rev. B {\bf 89}, 144502 (2014).
\bibitem{Xie20182} T. Xie, Y. Wei, D. Gong, T. Fennell, U. Stuhr, R. Kajimoto, K. Ikeuchi, S. Li, J. Hu, and H. Luo, Phys. Rev. Lett. {\bf 120}, 267003 (2018).
\bibitem{Xie2020} T. Xie, C. Liu, F. Bourdarot, L. -P. Regnault, S. Li, and H. Luo, Phys. Rev. Research {\bf 2}, 022018(R) (2020).
\bibitem{Supplementary} See Supplemental Materials for the sample information, polarized analysis and band caclulations, which includes Refs. [48 - 63].
\bibitem{ychen2011} Y. Chen, X. Lu, M. Wang, H. Luo, and S. Li, Supercond. Sci. Technol. {\bf 24}, 065004 (2011).
\bibitem{hqluo2012} H. Luo, R. Zhang, M. Laver, Z. Yamani, M. Wang, X. Lu, M. Wang, Y. Chen, S. Li, S. Chang, J. W. Lynn, and P. Dai, Phys. Rev. Lett. {\bf 108}, 247002 (2012).
\bibitem{xylu2013} X. Lu, H. Gretarsson, R. Zhang, X. Liu, H. Luo, W. Tian, M. Laver, Z. Yamani, Y. -J. Kim, A. H. Nevidomskyy, Q. Si, and P. Dai, Phys. Rev. Lett. {\bf 110}, 257001 (2013).
\bibitem{rzhang2015} R. Zhang, D. Gong, X. Lu, S. Li, M. Laver, C. Niedermayer, S. Danilkin, G. Deng, P. Dai, and H. Luo, Phys. Rev. B {\bf 91}, 094506 (2015).
\bibitem{txie2017} T. Xie, D. Gong, W. Zhang, Y. Gu, Z. Huesges, D. Chen, Y. Liu, L. Hao, S. Meng, Z. Lu, S. Li, and H. Luo, Supercond. Sci. Technol. {\bf 30}, 095002 (2017).
\bibitem{pgiannozzi2017} P. Giannozzi, O. Andreussi, T. Brumme {\it et al.}, J. Phys.: Condensed Matter {\bf 29}, 465901 (2017).
\bibitem{jpperdew1996} J. P. Perdew, K. Burke, and M. Ernzerhof, Phys. Rev. Lett. {\bf 77}, 3865 (1996).
\bibitem{lbellaiche2000} L. Bellaiche, D. Vanderbilt, Phys. Rev. B {\bf 61}, 7877(2000).
\bibitem{hjmonkhorst1976} H. J. Monkhorst and J. D. Pack, Phys. Rev. B {\bf 13}, 5188 (1976).
\bibitem{jdpackt1977}  J. D. Pack and H. J. Monkhorst, Phys. Rev. B {\bf 16}, 1748 (1977).
\bibitem{rfletcher1970} R. Fletcher, The Computer Journal {\bf 13}, 317 (1970).
\bibitem{dgoldfarb1970} D. Goldfarb, Math.Comp. {\bf 24}, 23 (1970).
\bibitem{dfshanno1970} D. F. Shanno, Math.Comp. {\bf 24}, 647 (1970).
\bibitem{mkawamura2019}  M. Kawamura, Comp. Phys. Commun. {\bf 239}, 197 (2019).
\bibitem{aghosh2021} A. Ghosh, S. Sen, H. Ghosh, Comp. Mater. Sci. 186, 109991(2021).
\bibitem{flochner2017} F. Lochner, F. Ahn, T. Hickel, and I. Eremin, Phys. Rev. B {\bf 96}, 094521 (2017).
\bibitem{Berna2005} E. Leli\`{e}vre-Berna, E. Bourgeat-Lami, P. Fouilloux {\it et al.}, Physica B {\bf 356}, 131 (2005).
\bibitem{Gong2018b} D. Gong, T. Xie, R. Zhang, J. Birk, C. Niedermayer, F. Han, S. H. Lapidus, P. Dai, S. Li, and H. Luo, Phys. Rev. B {\bf 98}, 014512 (2018).
\bibitem{gshirane2004} G. Shirane, S. M. Shapiro, J. M. Tranquada, {\it Neutron Scattering with a Triple-Axis Spectrometer}, Cambridge University Press, P23 - P46 and P170-P172 (2004).
\bibitem{Harriger2009} L. W. Harriger, A. Schneidewind, S. Li, J. Zhao, Z. Li, W. Lu, X. Dong, F. Zhou, Z. Zhao, J. Hu, and P. Dai, Phys. Rev. Lett. {\bf 103}, 087005 (2009).
\bibitem{Qureshi2012} N. Qureshi, P. Steffens, S. Wurmehl, S. Aswartham, B. B\"{u}chner, M. Braden, Phys. Rev. B {\bf 86}, 060410(R) (2012).
\bibitem{CWang2013} C. Wang, R. Zhang, F. Wang, H. Luo, L. P. Regnault, P. Dai, and Y. Li, Phys. Rev. X {\bf 3}, 041036 (2013).
\bibitem{RYu2020} R. Yu, H. Hu, E. M. Nica, J. -X. Zhu, Q. Si, Front. Phys. {\bf 9}, 578347 (2021).
\bibitem{Chen2021} L. Chen, T. T. Han, C. Cai, Z. G. Wang, Y. D. Wang, Z. M. Xin, and Y. Zhang, Phys. Rev. B {\bf 104}, L060502 (2021).
\bibitem{Zhang2018} W. -L. Zhang, W. R. Meier, T. Kong, P. C. Canfield, and G. Blumberg, Phys. Rev. B {\bf 98}, 140501(R) (2018).
\bibitem{Fernandes2012} R. M. Fernandes and J. Schmalian, Supercond. Sci. and Techn. {\bf 25}, 084005 (2012).
\bibitem{Liu2020} P. Liu, M. L. Klemm, L. Tian, X. Lu, Y. Song, D. W. Tam, K. Schmalzl, J. T. Park, Y. Li, G. Tan, Y. Su, F. Bourdarot, Y. Zhao, J. W. Lynn, R. J. Birgeneau and P. Dai, Nat. Commun. {\bf 11}, 5728 (2020).
\bibitem{RawData} Y. Li, F. Bourdarot, P. Bourges, G. He, C. Liu, H. Luo, Y. Sidis, X. Wang, and T. Xie (2020). Institut Laue-Langevin (ILL), doi:10.5291/ILL-DATA.CRG-2609.

\end{thebibliography}

\begin{thebibliography}{}
\bibitem{xie2018s1} T. Xie, Y. Wei, D. Gong, T. Fennell, U. Stuhr, R. Kajimoto, K. Ikeuchi, S. Li, J. Hu, and H. Luo, Phys. Rev. Lett. {\bf 120}, 267003 (2018).
\bibitem{ychen2011s} Y. Chen, X. Lu, M. Wang, H. Luo, and S. Li, Supercond. Sci. Technol. {\bf 24}, 065004 (2011).
\bibitem{hqluo2012s} H. Luo, R. Zhang, M. Laver, Z. Yamani, M. Wang, X. Lu, M. Wang, Y. Chen, S. Li, S. Chang, J. W. Lynn, and P. Dai, Phys. Rev. Lett. {\bf 108}, 247002 (2012).
\bibitem{xylu2013s} X. Lu, H. Gretarsson, R. Zhang {\it et al.}, Phys. Rev. Lett. {\bf 110}, 257001 (2013).
\bibitem{rzhang2015s} R. Zhang, D. Gong, X. Lu, S. Li, M. Laver, C. Niedermayer, S. Danilkin, G. Deng, P. Dai, and H. Luo, Phys. Rev. B {\bf 91}, 094506 (2015).
\bibitem{txie2017s} T. Xie, D. Gong, W. Zhang, Y. Gu, Z. Huesges, D. Chen, Y. Liu, L. Hao, S. Meng, Z. Lu, S. Li, and H. Luo, Supercond. Sci. Technol. {\bf 30}, 095002 (2017).
\bibitem{LELIEVREBERNA2005s}E. Leli\`{e}vre-Berna, E. Bourgeat-Lami, P. Fouilloux {\it et al.},  Physica B {\bf 356}, 131 (2005).
\bibitem{neutron_book_tripleaxiss} G. Shirane, S. M. Shapiro, J. M. Tranquada, Neutron Scattering with a Triple-Axis Spectrometer, Cambridge University Press, P117-P121 (2004).
\bibitem{lipscombe2010s} O. J. Lipscombe, L. W. Harriger, P. G. Freeman {\it et al.},  Phys. Rev. B {\bf 82}, 064515 (2010).
\bibitem{liu2012s} M. Liu, C. Lester, J. Kulda, X. Lu, H. Luo, M. Wang, S. M. Hayden, and P. Dai, Phys. Rev. B {\bf 85}, 214516 (2012).
\bibitem{steffens2013s} P. Steffens, C. H. Lee, N. Qureshi, K. Kihou, A. Iyo, H. Eisaki, and M. Braden, Phys. Rev. Lett. {\bf 110}, 137001 (2013).
\bibitem{zhang2014s} C. Zhang, Y. Song, L.-P. Regnault, Y. Su, M. Enderle, J. Kulda, G. Tan, Z. C. Sims, T. Egami, Q. Si, and P. Dai, Phys. Rev. B {\bf 90} 140502(R) (2014).
\bibitem{luo2013s} H. Luo, M. Wang, C. Zhang, X. Lu, L.-P. Regnault, R. Zhang, S. Li, J. Hu, and P. Dai, Phys. Rev. Lett. {\bf 111}, 107006 (2013).
\bibitem{qureshi2014s} N. Qureshi, C. H. Lee, K. Kihou, K. Schmalzl, P. Steffens, and M. Braden, Phys. Rev. B {\bf 90}, 100502 (2014).
\bibitem{song2016s} Y. Song, H. Man, R. Zhang, X. Lu, C. Zhang, M. Wang, G. Tan, L.-P. Regnault, Y. Su, J. Kang, R. M. Fernandes, and P. Dai, Phys. Rev. B {\bf 94}, 214516 (2016).
\bibitem{song2017s} Y. Song, W. Wang, C. Zhang, Y. Gu, X. Lu, G. Tan, Y. Su, F. Bourdarot, A. D. Christianson, S. Li, and P. Dai, Phys. Rev. B {\bf 96}, 184512 (2017).
\bibitem{hu2017s} D. Hu, W. Zhang, Y. Wei, B. Roessli, M. Skoulatos, L. P. Regnault, G. Chen, Y. Song, H. Luo, S. Li, and P. Dai, Phys. Rev. B {\bf 96}, 180503 (2017).
\bibitem{zhang2013s} C. Zhang, M. Liu, Y. Su, L.-P. Regnault, M. Wang, G. Tan, Th. Br¨¹ckel, T. Egami, and P. Dai, Phys. Rev. B {\bf 87}, 081101(R) (2013).
\bibitem{qureshi2012s} N. Qureshi, P. Steffens, S. Wurmehl, S. Aswartham, B. B¨¹chner, and M. Braden, Phys. Rev. B {\bf 86}, 060410(R) (2012).
\bibitem{xie2018s2} T. Xie, D. Gong, H. Ghosh, A. Ghosh, M. Soda, T. Masuda, S. Itoh, F. Bourdarot, L.-P. Regnault, S. Danilkin, S. Li, and H. Luo, Phys. Rev. Lett. {\bf 120}, 137001 (2018).
\bibitem{xie2020s} T. Xie, C. Liu, F. Bourdarot, L.-P. Regnault, S. Li, and H. Luo, Phys. Rev. Research {\bf 2}, 022018 (2020).
\bibitem{giannozzi2017s} P. Giannozzi, O. Andreussi, T. Brumme, {\it et al.}, J. Phys.: Condens. Matter {\bf 29}, 465901 (2017).
\bibitem{perdew1996s} J. P. Perdew, K. Burke, and M. Ernzerhof, Phys. Rev. Lett. {\bf 77}, 3865 (1996).
\bibitem{bellaiche2000s} L. Bellaiche and D. Vanderbilt, Phys. Rev. B {\bf 61}, 7877 (2000).
\bibitem{monkhorst1976s} H. J. Monkhorst and J. D. Pack, Phys. Rev. B {\bf 13}, 5188 (1976).
\bibitem{pack1977s} J. D. Pack and H. J. Monkhorst, Phys. Rev. B {\bf 16}, 1748 (1977).
\bibitem{fletcher1970s} R. Fletcher, The Computer Journal {\bf 13}, 317 (1970).
\bibitem{goldfarb1970s} D. Goldfarb, Math. Comp. {\bf 24}, 23 (1970).
\bibitem{shanno1970s} D. F. Shanno, Math. Comp. {\bf 24}, 647 (1970).
\bibitem{kawamura2019s} M. Kawamura, Comp. Phys. Commun. {\bf 239}, 197 (2019).
\bibitem{ghosh2021s} A. Ghosh, S. Sen, and H. Ghosh, Comp. Mater. Sci. {\bf 186}, 109991 (2021).
\bibitem{lochner2017s} F. Lochner, F. Ahn, T. Hickel, and I. Eremin, Phys. Rev. B {\bf 96}, 094521 (2017).
\end{thebibliography}
\end{document}